\documentclass[12pt]{article}

\usepackage{color}
\usepackage[pagebackref]{hyperref}
\usepackage{subcaption}
\usepackage{algorithm}
\usepackage{algpseudocode}
\usepackage{amsbsy}
\usepackage{amsfonts}
\usepackage{amsmath}
\usepackage{amssymb}
\usepackage{algpseudocode}
\usepackage{mathtools}
\usepackage{amscd,amsxtra}
\usepackage{dsfont}
\usepackage{url}
\usepackage{bbm}
\usepackage{textcomp}
\usepackage{float}
\usepackage{natbib}

\DeclareMathOperator*{\argmax}{argmax}

\def\spacingset#1{\renewcommand{\baselinestretch}%
{#1}\small\normalsize} \spacingset{1}

\begin{document}
  \title{\bf Analysis and Optimization of Seismic Monitoring Networks with Bayesian Optimal Experiment Design}  
\author{Jake Callahan \thanks{jpcalla@sandia.gov
    }\hspace{.2cm}\\
    Sandia National Laboratories, Livermore, CA 94550\\
    Program in Applied Mathematics, University of Arizona, Tucson, AZ 85719\\
    Kevin Monogue\thanks{kcmonogue@gmail.com} \\
    Stanford University, Stanford, CA 94305\\
    Ruben Villarreal \thanks{rubvill@sandia.gov} \\
    Sandia National Laboratories, Livermore, CA 94550\\
    and \\
    Tommie Catanach \thanks{tacatan@sandia.gov} \\
    Sandia National Laboratories, Livermore, CA 94550}
    \date{}
  \maketitle
\bigskip
\begin{abstract}
    Monitoring networks increasingly aim to assimilate data from a large number of diverse sensors covering many sensing modalities. Bayesian optimal experimental design (OED) seeks to identify data, sensor configurations, or experiments which can optimally reduce uncertainty and hence increase the performance of a monitoring network. Information theory guides OED by formulating the choice of experiment or sensor placement as an optimization problem that maximizes the expected information gain (EIG) about quantities of interest given prior knowledge and models of expected observation data. Therefore, within the context of seismo-acoustic monitoring, we can use Bayesian OED to configure sensor networks by choosing sensor locations, types, and fidelity in order to improve our ability to identify and locate seismic sources.  In this work, we develop the framework necessary to use Bayesian OED to optimize a sensor network's ability to locate seismic events from arrival time data of detected seismic phases at the regional-scale. Bayesian OED requires four elements:
\begin{enumerate}
\item A likelihood function that describes the distribution of detection and travel time data from the sensor network,

\item A Bayesian solver that uses a prior and likelihood to identify the posterior distribution of seismic events given the data,

\item An algorithm to compute EIG about seismic events over a dataset of hypothetical prior events,

\item An optimizer that finds a sensor network which maximizes EIG.
\end{enumerate}
Once we have developed this framework, we explore many relevant questions to monitoring such as: how to trade off sensor fidelity and earth model uncertainty; how sensor types, number, and locations influence uncertainty; and how prior models and constraints influence sensor placement.
\end{abstract}

\noindent%
\textit{Keywords:} Bayesian inference, Optimal experiment design, Statistical methods, probability distributions, seismic noise, seismic instruments, statistical seismology, earthquake source observations, earthquake interaction, forecasting, and prediction, earthquake monitoring and test-ban treaty verification

\section{Introduction}

Seismo-acoustic monitoring networks are central to detecting and locating earthquakes, explosions or other seismic sources. In order to improve monitoring capabilities, network designers may incorporate new sensors or data types into the network to reduce detection thresholds or improve estimate uncertainties for quantities of interest (QoIs) like location, magnitude, and depth. To estimate a QoI, modern processing algorithms often employ Bayesian inference because it provides rigorous uncertainty quantification to support decision making \citep{arora2013net,myers2007bayesian}. Therefore, when designing or analysing a monitoring network, we approach it from the philosophy of Bayesian Optimal Experimental Design (OED) \citep{krause2008near,huan2013simulation}. Within this framework, we optimize sensors of a monitoring network to reduce uncertainty about QoIs under different conditions described by a prior distribution. Therefore, with Bayesian OED we not only design an effective monitoring network, but also get an understanding of the expected performance of that network under the specified conditions. While the target application of this research is explosion monitoring, the experimental design framework we have developed applies to arbitrary seismic sources. Therefore, this framework may support other applications of seismic networks such as earthquake seismology, earthquake or tsunami early warning, or exploration geophysics.

Optimal experimental design has been a recent active area of research in many areas of seismology including Early Warning, Seismic Source Inversion, Tomography, and Structural Health Monitoring.  Typically these studies have focused on network design in terms of the number and location of sensors \citep{bose2022loss, papadimitriou2005optimal, yuen2015efficient, yang2022optimal, guest2011standard, an2018sensitivity, toledo2020optimized, bloem2020experimental}, although some work has also explored different sensor types \citep{yuen2015efficient}. For linear inverse problems, or those that can be linearly approximated, the alphabetic optimality criteria like D-optimal design are often used as an objective \citep{koval2021optimal, coles2011efficient, steinberg2003optimal, burmin2019optimal, bloem2020experimental}. For non-linear inverse problems, Bayesian methods have become popular, leveraging information-based metrics such as entropy-based design or mutual information \citep{long2015fast, yang2022optimal, maurer2010recent, bloem2020experimental}. However, because of the computational cost of these methods, many different approximations methods have been explored to speed up estimating the objective \citep{long2015fast, maurer2010recent, coles2012toward}. Finally, work has also explored different optimizers to explore the configuration space of networks ranging from genetic algorithms to gradient-based methods \citep{bose2022loss, oth2010evaluation, papadimitriou2005optimal, curtis2004deterministic, guest2011standard, toledo2020optimized}. All these considerations lead to a tradeoff between computational tractability and accuracy which has started to be explored.

Our work adds to this body of recent research through the following contributions:
\begin{enumerate}
    \item Presenting a holistic treatment of uncertainty (e.g. model error, measurement error, sensor correlation, etc.) for the Bayesian OED problem for seismic monitoring
    \item Studying OED in a broader context than just sensor placement e.g., relative trade-offs in model refinement vs data fidelity.
    \item Introducing a Bayesian optimization algorithm to efficently optimize the sensor network.
    \item Releasing a computationally efficient grid method for fully Bayesian OED leveraging HPC that can be widely used for seismo-acoustic monitoring network design and analysis. 
\end{enumerate}

This work quantifies the sensitivity of a seismo-acoustic monitoring network for inferring the location and magnitude of seismic sources that include shallow earthquakes and explosions either on the surface or underground. We then present a Bayesian OED algorithm to improve the monitoring network sensitivity by optimizing the location of ground motion sensors. Our computational approach combines information and Bayesian probability theory to quantify and optimize the sensitivity of our sensor network by estimating the information gain Bayesian inference provides about QoIs. This approach includes four distinct analysis stages:
\begin{enumerate}
\item Build the likelihood function to estimate the probability of data, given a seismic event.
\item Solve the Bayesian inference problem for locating events given data (e.g., solve for the posterior).
\item Estimate seismic source location sensitivity through measuring the expected information gain, with a sensor network.
\item Optimize the seismic monitoring network to improve the expected information gain over seismic source events.
\end{enumerate}

We use observational data from the U.S. Transportable Array \citep{usarray} and physics-based models to build the Bayesian likelihood functions. 
These likelihood functions incorporate many sources of uncertainty and model the behavior of the seismic sensor network. 
This model defines how well Bayesian inference can assimilate sensor data to locate seismic sources. For our sensor data, we consider spatially correlated travel-times for seismic phase arrivals detected at our sensor network. We make generally justifiable assumptions on our uncertainty models that match properties of the datasets that we consider. While we present our method using seismic P-wave arrivals, our approach is flexible to any event or signature data because it only requires that we can construct likelihood functions.

We study how the optimized sensor configuration and network sensitivity change under different design conditions and uncertainty models. We present the dependence of our results over prior sensor distribution, sensor number, and sensor fidelity. This analysis provides a framework that we can later extend to optimize sensor networks that measure other natural and explosion signatures (e.g., electromagnetic or infrasound signals), which supports a more comprehensive need for multi-phenomenology explosion monitoring (e.g., \cite{Carmichael2020_1,arrowsmith2020event}). 
This framework explores an alternative approach to existing tools, like Sandia National Laboratory's NetMOD 
\citep{osti_1337571}, with the aim to provide a highly flexible and rigorous framework for analysing and optimizing monitoring networks. This rigor is justified through our usage of Bayesian probability theory and uncertainty quantification.

In Section \ref{sec:Bayes}, we describe the Bayesian inference and optimal experimental design problems in general. In Section \ref{sec:BSM} we describe the specifics of a Bayesian inference problem to identify the location and magnitude of seismic sources, using records of their P-phase arrivals at distributed receivers and demonstrate how to build the likelihood models from these data. Next, in Section \ref{sec:comp} we describe the algorithms used for solving the Bayesian OED problem. Finally, in Section \ref{sec:results} we will describe several experiments that demonstrate the utility of this approach and identify areas for further exploration. Section \ref{sec:conclusion} concludes with discussion and future work.

\section{Bayesian Methods}
\label{sec:Bayes}

\subsection{Bayesian Inference}
Bayesian probability theory provides a rigorous methodology to quantify and update uncertainty about beliefs \citep{beck2010bayesian, jaynes2003probability, gelman1995bayesian}. Within this framework, uncertainty is represented using probability distributions. Therefore, within the Bayesian paradigm, probability distributions represent uncertainty about beliefs and not necessarily intrinsic stochasticity and thus are not directly tied to randomness. Uncertainty comes from both epistemic sources, when it represents a lack of knowledge about learnable phenomena (ignorance), or aleatory sources, when it represents uncertainty about inherently unknowable randomness (unresolvable uncertainty for the observer). The Bayesian perspective describes both of these sources of uncertainty using a probability distribution. Therefore, the Bayesian framework can helpfully incorporate modeling error, measurement error, and parametric uncertainty.

As data or other sources of information become available, an observer can integrate these information into a new probability distributions to update the observer's beliefs. When the observer makes predictions, they include the uncertainty represented by these probability distributions in these predictions. The rules of Bayesian probability provide a rigorous logic for updating and propagating uncertainty just as binary logic provides rules for working with statements that are true or false.

\begin{figure}
\centering
\includegraphics[scale=0.4]{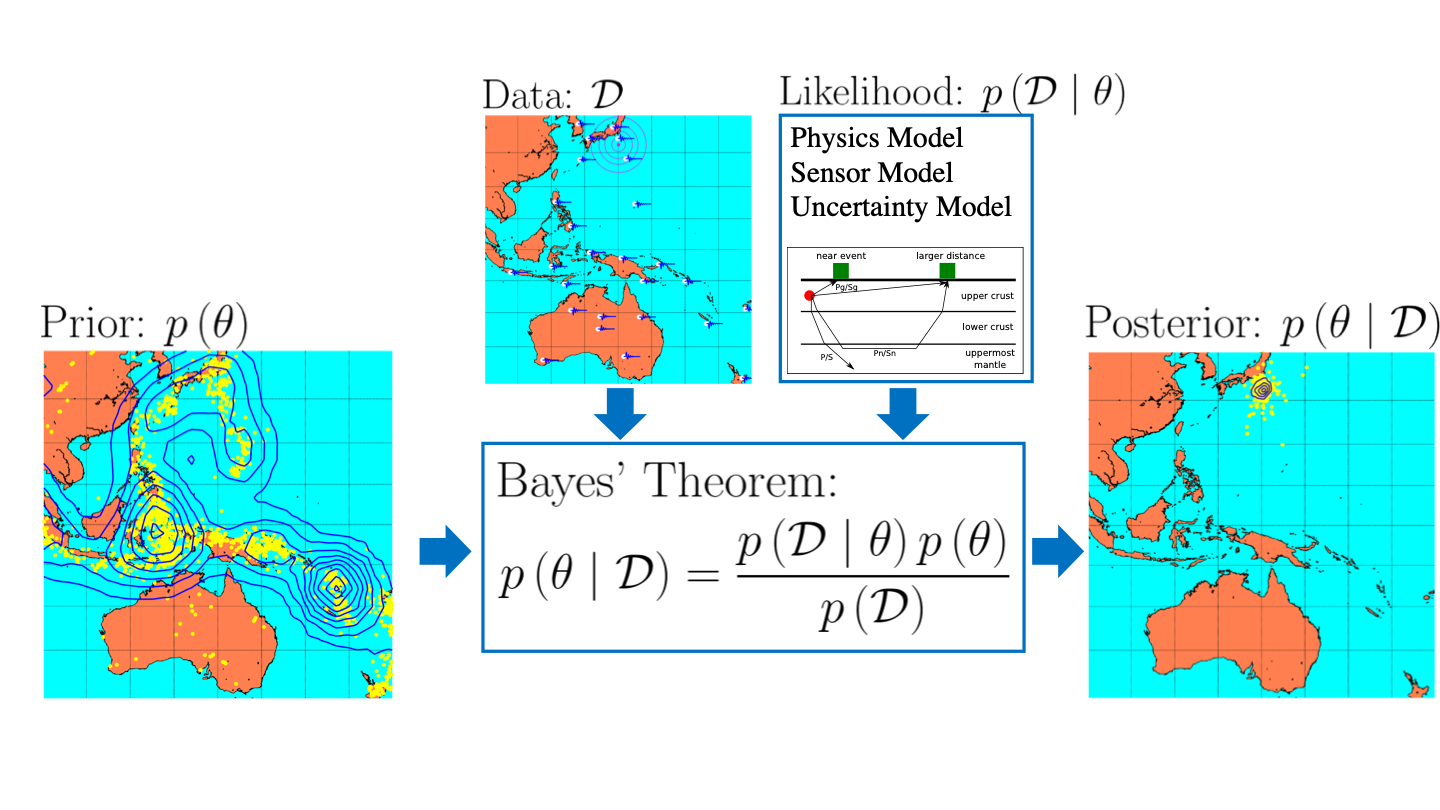}
\caption{Illustration of a Bayesian inference process for seismic source location. Bayesian inference begins with a prior distribution for different earthquake locations $\theta$, shown by the contour lines on the leftmost figure. As an observer collects data, they use a likelihood function model to quantify the probability of observing that data, given that an earthquake occurs at a specific location. The observer constructs this likelihood model from physical models of seismic wave propagation, models of the sensors that detect seismic signals, and models of uncertainty (e.g. background noise modeling errors, etc.). The observer then applies Bayes' theorem to update the prior to assimilate this new information. The posterior distribution, shown by the contour lines in the rightmost image, then quantifies the probability that the seismic source has location  $\theta$, given the data. }
\label{fig:bayes}
\end{figure}

The process of updating beliefs using data are known as Bayesian inference. Figure \ref{fig:bayes} illustrates Bayesian inference applied to a hypothetical seismic location problem (see \cite{myers2007bayesian, arora2013net} for some detailed applications of Bayesian inference to locating seismic sources). Bayesian inference begins by expressing prior beliefs about parameters of interest $\theta$. For example, within the context of identifying characteristics of a seismic event, these beliefs may represent prior knowledge about the distribution of earthquake magnitudes or their locations, e.g., source proximity to lithospheric faults. As an observer gathers data $\mathcal{D}$ and other information, they update the prior distribution using the rules of probability. This updated, or posterior, distribution $p \left (\theta \mid \mathcal{D} \right )$ now quantifies the likelihood of the source location given the data. The observer performs this update using a likelihood function to describe the probability of the data given an event hypothesis, i.e., $p \left ( \mathcal{D} \mid \theta \right ) $. An observer constructs such a likelihood function from a probabilistic forward model of the data observed given a set of source parameters. This means that that the likelihood can equally be used to construct a generative model of the data given the source parameters. The likelihood function assumes specified source parameters that describe the seismic source and then uses physical models, sensor models, and models of background signals and noise to map this source description to plausible sensor data. As an example, if the arrival time of a seismic phase at a seismometer constitutes observed data, and event parameters describe the location and origin time of an earthquake, then the likelihood uses an earth structure model to predict uncertainty in the arrival time of a seismic phase from the source to any receivers. The model of the travel time could include (predicted) earth model uncertainty and measurement uncertainty on the sensor.

Once an observer constructs a likelihood function, they can easily construct the posterior distribution on events given data. This construction is an application of Bayes' Theorem:

\begin{equation}
\label{eq:bayes}
p \left (\theta\mid \mathcal{D} \right ) = \frac{p \left ( \mathcal{D} \mid \theta \right ) p \left ( \theta \right )}{p\left ( \mathcal{D} \right )}
\end{equation}

We emphasize that the probability terms in Equation \ref{eq:bayes} can be either probabilities when $\theta$ is a discrete random variable or a probability density when $\theta$ is continuous.  Equation \ref{eq:bayes} provides the foundational statement of belief about uncertainties in the model and the machinery to update these beliefs as new information becomes available. In practice, solving for the updated Bayesian posterior requires approximate computational methods since the posterior may not have an analytical expression. Common approaches generate samples representing draws from the posterior distribution and can estimate QoIs. Examples of these methods include importance sampling using Monte Carlo, Quasi Monte Carlo, meshing, and Markov Chain Monte Carlo \citep{brooks2011handbook, owen2013monte}.

\subsection{Bayesian OED}
To quantify network performance, we require a measure of how much belief changes due to inference on observed data. This is a measure of the sensor data's utility that defines the objective for experimental design. One measure that is commonly used in information theory is the Kullback-Leibler divergence: 

\begin{equation}
\label{eq:KLdiv}
\text{KL} \left [p \left (\theta\mid \mathcal{D} \right ) \mid \mid  p \left ( \theta \right ) \right ] = \int p \left (\theta\mid \mathcal{D} \right ) \log \frac{p \left (\theta\mid \mathcal{D} \right )}{p \left (\theta\right )} d\theta
\end{equation}

\noindent The KL divergence in Equation \ref{eq:KLdiv} measures how many units of information (bits for $\log_2$ or nats for $\ln$) are needed to specify a change in the distribution from $p \left (\theta\right )$ to $p \left (\theta\mid \mathcal{D} \right )$. These units are related to the efficiency of encoding a probability distribution (see \cite{mackay2003information} for discussion). A KL divergence of $0$ means that the distributions are the same up to certain technical details. The KL divergence is always non-negative and as it increases from zero, Equation \ref{eq:KLdiv} implies that the distributions increasingly differ. A relatively large KL divergence therefore indicates that the data was very informative, and the prior and posterior are measurably distinct.

The Bayesian optimal experimental design (OED) problem is built upon the concepts of Bayesian probability and information theory \citep{huan2013simulation, ginebra2007measure, lindley1956measure}. Bayesian OED assumes that the observer applies Bayes' Theorem (i.e., that they are a Bayesian agent) to select a sensor configuration $\mathcal{S}$ that maximizes utility; we term $\mathcal{S}$ as the ``experiment.''  Because Bayesian inference is the optimal way to assimilate information it provides, Bayesian OED defines the best case scenario for extracting information from the sensor network. The Bayesian agent optimizes a utility function that depends on the posterior. In this research, the Bayesian agent maximizes the expected information gain (EIG) from the prior to the posterior, in the view of the posterior. Notationally, the expectation $E_{\mathcal{D} \mid \mathcal{S}}$ indicates that the observer computes the expectation with respect to the prior distribution of hypothetical data from the experiment given by $p \left (  \mathcal{D} \mid \mathcal{S} \right)$.  The expected information gain for a specific experimental configuration is (from Equation \ref{eq:KLdiv}):

\begin{align}
\label{eq:expectKL}
\mathcal{I} \left ( \mathcal{S}  \right )
&= \text{E}_{\mathcal{D} \mid \mathcal{S}} \left [ \text{KL} \left [p \left (\theta \mid \mathcal{D}, \mathcal{S} \right ) \mid \mid  p \left ( \theta \right ) \right ]\right ] \nonumber \\ 
 &=\int p \left ( \mathcal{D} \mid \mathcal{S} \right) \int p \left (\theta\mid \mathcal{D},   \mathcal{S}  \right ) \log \frac{p \left (\theta\mid \mathcal{D},  \mathcal{S}  \right )}{p \left (\theta\right )} d\theta d\mathcal{D}
\end{align}

The outer integral in Equation \ref{eq:expectKL} is the expectation over the hypothetical data from the experiment, while the inner integral computes the KL divergence given a realization of the hypothetical data. To compute the EIG in practice, we express $p \left (  \mathcal{D} \mid \mathcal{S} \right)$ as the marginal distribution 
\[
p \left (  \mathcal{D} \mid \mathcal{S} \right) = \int p \left (  \mathcal{D} \mid \theta^{\prime}, \mathcal{S} \right) p \left ( \theta^{\prime} \right ) d\theta^{\prime}.
\] because the likelihood of the data is often only implicitly known by definition of the likelihood and prior over parameters $\theta^{\prime}$. Note here that we have assumed that $p(\theta)$ is a proper density. We then draw samples from the marginal distribution by first sampling the prior, and then sampling the data according to the likelihood. These samples allow us to compute the outer expectation.

We now maximize  $\mathcal{I} \left ( \mathcal{S}  \right )$ to estimate the best experimental design $\mathcal{S}^*$ from $\mathcal{S} \in \mathbb{S}$, where $\mathbb{S}$ is the set of possible designs under consideration:

\begin{equation}
\label{eq:optExpConfig}
\mathcal{S}^* = \argmax_{\mathcal{S} \in \mathbb{S}} \mathcal{I} \left ( \mathcal{S}  \right )
\end{equation}

\noindent This optimization is generalizable to include constraints that include, for example, a sensor budget or constraints on sensor locations through methods like Lagrange multipliers or nonlinear programming. Further, while we have formulated this problem through maximizing the EIG for the posterior, we could more specifically optimize EIG about a specific quantity of interest derived from its parameters.

Solving this optimization problem is challenging because it requires solving many Bayesian inference problems for many hypothetical realizations of data from many hypothetical sensor configurations. This nested complexity means that significant care must be taken to make this approach computationally tractable.

\subsection{Bayesian Optimization}
\label{sec:bayesopt}
\begin{figure*}
\centering
\centerline{\includegraphics[scale=0.3]{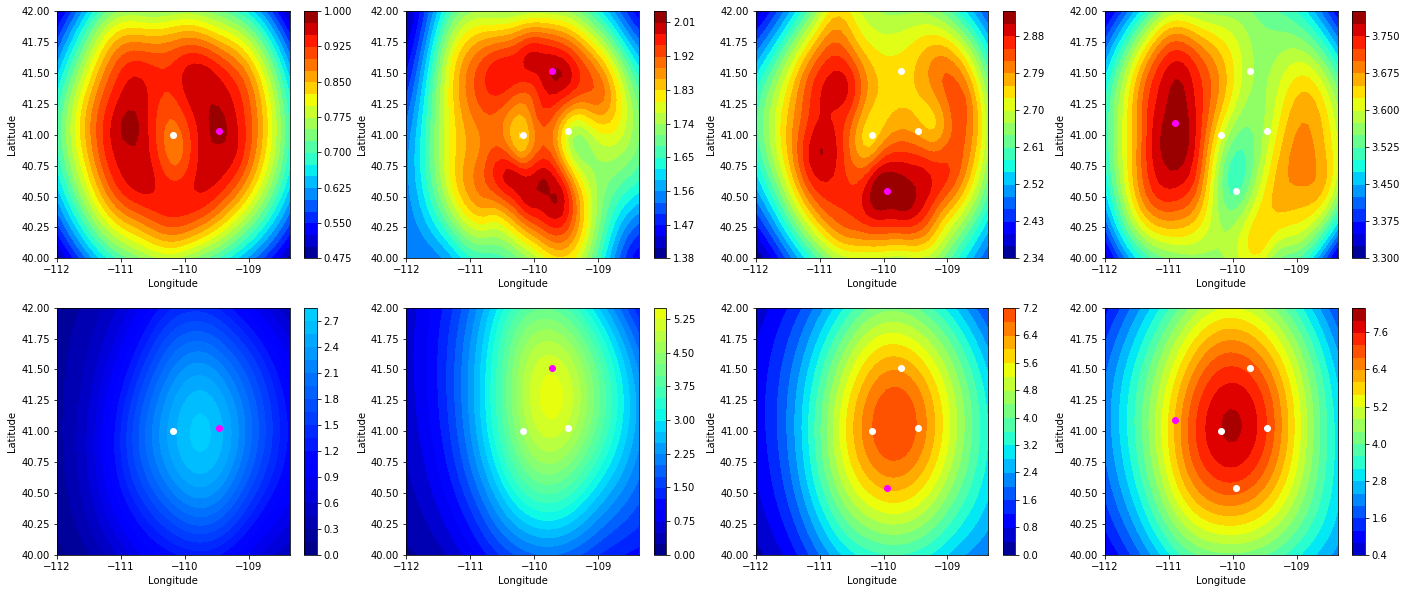}}
\caption{Sensor locations that optimize a sensor network configuration, according to Equation \ref{eq:optExpConfig}. 
The top row displays heat maps showing the optimization surface $\mathcal{I}(S_{n+1})$, where $S_{n+1} = \{S^*_n, \mathcal{L}_{n+1}\}$, $S^*_n$ is the set of the first $n$ fixed sensor locations (white circular markers), and $\mathcal{L}_{n+1}$ is the location of the $(n+1)$\textsuperscript{th} sensor (magenta circular markers) that maximizes $\mathcal{I}(S_{n+1})$. The warmer the color indicates that adding a sensor at that location is better.
The bottom row displays the EIG about all events in the domain about the location of a shallow, low-magnitude seismic source at that specific latitude and longitude, i.e., it displays $\mathcal{I}(S^*_{n+1} \mid \theta' = [\mathcal{L},x,m])$ for all $\mathcal{L}$ in the domain (see Equation \ref{eq:thetaEIG}).}
\label{fig:greedy_opt}
\end{figure*}

Greedy optimization algorithms provide an effective computational solution to sequentially place sensors in OED and for other network optimization problems 
\citep{krause2008near, Carmichael2021}. Such greedy optimization involves sequentially adding sensors one at a time so that the optimization problem at a particular iteration is low dimensional, and therefore only requires updating the location of that particular sensor. During iteration, the algorithm computes the EIG as an average over all possible source locations specified by the prior, and then computes an optimal location. Figure \ref{fig:greedy_opt} illustrates the process of adding sensors one-by-one. 
 The optimization surfaces, shown in the top row of the figure, start out fairly symmetric with multiple optima when there are few sensors. However, as more sensors are added these symmetries are broken so there is a unique optimal location of the next sensor. The bottom row illustrates how the EIG increases with sensor density, particularly about sources that are near several sensors.

We concede that the true, optimal sensor network configuration requires that we compute sensor location solutions all at once. However, greedy optimization often does reasonably well with a significantly reduced computational cost. In fact, suboptimal bounds exist for certain classes of optimization problems approximated using greedy methods, such as the class of submodular functions. At a high level, submodular utility functions exhibit diminishing returns with each iteration, that is, adding a sensor to a smaller network yields higher gains that adding a sensor to a larger network. The EIG objective for Bayesian OED is submodular when the sensors are conditionally independent given the event description, i.e., when
\[
p(\mathcal{D}_i, \mathcal{D}_j \mid \theta) = p(\mathcal{D}_i \mid \theta)p(\mathcal{D}_j \mid \theta),
\]
where $\mathcal{D}_i$ and $\mathcal{D}_j$ are data generated by sensors $i$ and $j$, respectively. If the utility function is submodular, then we can show that the greedy optimum will be near-optimal, meaning that there is a multiplicative approximation guarantee of $\left (1 - 1/e \right )$. Therefore, the greedy optimum is guaranteed to be within approximately 63\% of the global optimum. This bound is loose in practice, and stronger assumptions on the problem structure may yield smaller departures from global optimality. We refer to  \cite{krause2008near} for details on submodular functions and greedy optimization. 

We use Bayesian optimization \citep{movckus1975bayesian} to greedily optimize sensor placement locations. This requires sampling the utility function to build a surrogate model of the optimization surface from the samples, such as a Gaussian Process model \cite{williams2006gaussian}. Using this surrogate model, we choose new points at which to evaluate the utility function according to an acquisition function. The choice of acquisition function determines how we balance the exploration of high uncertainty regions of the parameter space, improving our surrogate model, with optimizing the existing surrogate to sample new points that will be close to the predicted optimum. The residual between the optimal solution and the sampled solution improves with iteration. Details of Bayesian optimization and descriptions of acquisition functions can be found in \cite{srinivas2009gaussian, frazier2018tutorial, jones1998bayesopt, picheny2013efficientopt}. We use the Python library \textsc{scikit-opt} \citep{head_tim_2020_4014775} to implement Bayesian optimization with a GP surrogate.

\section{Bayesian Seismic Monitoring}
\label{sec:BSM}
\subsection{General Approach}
As introduced in Figure \ref{fig:bayes}, Bayesian inference for seismic monitoring requires constructing a likelihood model $p \left ( \mathcal{D} \mid \theta,  \mathcal{S} \right )$ that quantifies the likelihood of the data given an event $\theta$ with a seismic sensor network configuration $\mathcal{S}$. We assume that a source can be sufficiently defined by a vector of its origin time, location, and magnitude $\theta = \left \{\text{Time}, \text{Lat}, \text{Long}, \text{Depth}, \text{Mag} \right \}$. Notationally, these  parameters are epicentral location $\mathcal{L} = \left \{ \text{Lat}, \text{Long} \right \}$, source depth $x$, event magnitude $m$, and origin time $t_o$. The network $\mathcal{S}$ consists of individual stations $\mathcal{S}_i$. Such stations may have heterogenous response or sampling features but here we assume they are homogenous. Therefore, station description is sufficiently described by $\mathcal{S}_i = \{\mathcal{S}^{\text{Loc}}_i\}$ where $\mathcal{S}^{\text{Loc}}$ is the station's location in latitude and longitude.

We limit our analysis to modeling arrivals of seismic phases from their sources and leave inclusion of waveform features to future research. Therefore, our data take the form of $\mathcal{D} = \{ \mathbb{D}, \mathbb{A} \}$, where $\mathbb{D}$ stores data about which stations detected different seismic phases and $\mathbb{A}$ stores information about the arrival times, $t_{ij}$, of the detected phases.

\begin{align}
\label{eq:dataD}
 \mathbb{D}_{ij} =
\begin{cases}
1  \,\, \text{if station $i$ detects phase $j$}\\
0  \,\, \text{if station $i$ does not detect phase $j$}\\
\end{cases}
\end{align}

\begin{align}
\label{eq:dataA}
 \mathbb{A}_{ij} =
\begin{cases}
t_{ij}  \,\, \text{if station $i$ detects phase $j$}\\
\emptyset  \,\, \text{if station $i$ does not detect phase $j$}\\
\end{cases}
\end{align}

Note that $\mathbb{A}_{ij} = \emptyset$ when no phase is detected since there is no arrival time to capture. We make the simplifying assumption that the likelihood of detection is independent of the origin time $t_o$. This assumption seems reasonable, but may not hold in cases that background noise is diurnally variable. Incorporating a time dependent background would not be difficult but is left to future work. We also assume that the priors are independent, but this likewise is easy to relax as needed. The posterior then is:

\begin{align}
p \left( \mathcal{L}, x, m, t_o \mid \mathbb{A}, \mathbb{D}, \mathcal{S}  \right ) &= \frac{p \left(  \mathbb{A}, \mathbb{D} \mid  \mathcal{L}, x, m, t_o, \mathcal{S}  \right ) p \left(   \mathcal{L}, x, m, t_o \right ) }{p \left(  \mathbb{A}, \mathbb{D} \mid \mathcal{S}  \right )}  \nonumber \\
&\propto p \left(  \mathbb{A} \mid  \mathcal{L}, x, m, t_o, \mathbb{D}, \mathcal{S}  \right ) p \left(\mathbb{D} \mid  \mathcal{L}, x, m, \mathcal{S}  \right ) p \left(   \mathcal{L} \right) p \left(  x \right) p \left(  m \right) p \left(  t_o \right) 
\end{align}

To construct the likelihood $p \left(\mathbb{D} \mid  \mathcal{L}, x, m, \mathcal{S}  \right )$ we must estimate the detection probability for a given arrival. When historic data are available, we can build a model for the detection of a phase at a station given an event at a specific location and with a specified magnitude as we discuss in Section \ref{sec:detect}. 

We consider two separate sources of uncertainty in the arrival time likelihood 
\[p \left(  \mathbb{A} \mid  \mathcal{L}, x, m, t_o, \mathbb{D}, \mathcal{S}  \right ):\] 
measurement noise and model prediction uncertainty.  We assume that the measurement noise distributions for each station are independent (which may not be true in situations where sensors are close, but provides a tractable simplifying assumption that is valid for sparse networks). For situations where measurement noise statistics are known a priori for a sensor and processing method, they be directly integrated into the likelihood model. Otherwise, the measurement noise model can be derived from data along with other assumptions that we will describe in  Section \ref{sec:measnoise}. However, when deriving measurement error models directly from data, the effect of modeling uncertainty must also be simultaneously accounted for. Unlike for measurement uncertainty, including correlated travel time errors across different stations in the likelihood function for modeling error is important. This correlation reflects that real Earth structure will likely be different than the modeled Earth structure and this discrepancy will induce correlated errors. We therefore model this uncertainty, and the correlation induced in the sensor network, by sampling a distribution of Earth models to estimate the distribution in arrival times as discussed in Section \ref{sec:modelnoise}.

\subsection{Detection Model}
\label{sec:detect}
\begin{figure*}
\centering
\centerline{\includegraphics[scale=0.3]{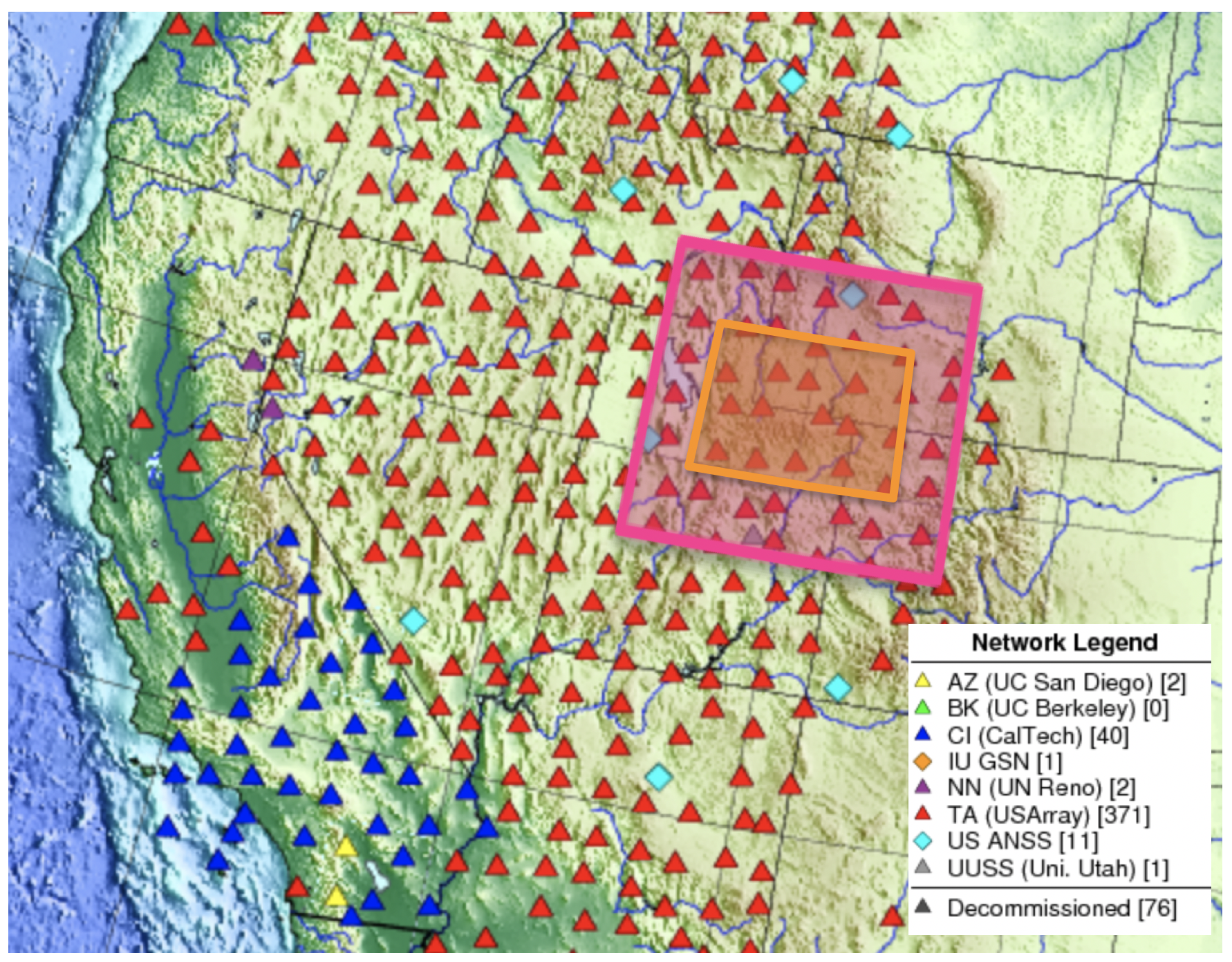}}
\caption{Map of Transportable Array stations from December 2007 (Map from \cite{stationmap}). The pink region indicates the region (Latitude $ \in [ 39^oN,  43^oN ]$, Longitude $ \in [ 113^oW,  107.36^oW ]$) that we use gathered sensor data that populates the parameters of the likelihood models. The orange region (Latitude $ \in [ 40^oN,  42^oN ]$, Longitude $ \in [ 112^oW,  109.36^oW ]$), corresponds to the monitoring region over which we build a sensor network \citep{usarray}.}
\label{fig:ta_map}
\end{figure*}

We use a catalog from the USArray Transportable Array experiment \citep{usarray} to build a detection model for seismic phases, specifically the first P arrival, i.e. arrivals labelled as P, Pg, and Pn in the catalog. Details of the modeling region can be seen in Figure \ref{fig:ta_map}. This model is similar to the logistic regression model used by NET-VISA \citep{arora2013net}. In principle this method can be followed for any monitoring region with existing sensors.
The USArray dataset was chosen because of the homogeneity of the sensor network and its uniform coverage for a region. Sensors were deployed in this region from August 2007 - August 2008 and during that time, 1089 events were registered on 45 stations. For these events, 11487 P arrivals were detected out of the 49005 potential detections, which corresponds to 23\% of potential P detections. Note we assume that every station had the potential to detect each event so the number of potential detections is the just the number of events multiplied by the number of stations. Of the 1089 events, 833 had estimated magnitudes. The minimum magnitude of the dataset was 0.51, maximum was 4.37, and median was 2.03.

Figure \ref{fig:detect_prob} shows the mean detection probability of seismic sources in our catalog, binned over magnitude and distance. We construct a logistic regression model using input features that include the distance between the event and the sensor (in degrees), the depth of the event, and the magnitude of the event. Our catalog data also included events with missing magnitude estimates.  We therefore used an additional indicator feature to reflect the absence of magnitude information in our source vector $\theta$. This feature is $1$ when the magnitude data are absent and $0$ when magnitude data are present. This feature helps us train using the data with missing magnitudes, which is critical for low magnitude events. When this model is used as part of the Bayesian OED framework, the magnitudes of hypothetical events will all be known so this indicator feature is always ignored after training.  As described previously, we assume that the detection probability for each station is conditionally independent, so the likelihood model becomes:

\begin{align}
p \left( \mathbb{D} \mid  \mathcal{L}, x, m, \mathcal{S}  \right ) &= \prod_{i} p \left( \mathbb{D}_{i} \mid \mathcal{L}, x, m, \mathcal{S}_i  \right )\\
p \left( \mathbb{D}_{i} \mid  \mathcal{L}, x, m, \mathcal{S}_i  \right ) &=
\begin{cases}
\frac{\exp \left ( \alpha Dist\left [  \mathcal{L}, \mathcal{S}_i \right ] + \beta x + \gamma m +\delta \right )}{1+\exp \left ( \alpha Dist\left [  \mathcal{L}, \mathcal{S}_i \right ] + \beta x + \gamma m +\delta \right )}, \,\, \text{if station $i$ detects the phase}\\
\frac{1}{1+\exp \left ( \alpha Dist\left [  \mathcal{L}, \mathcal{S}_i \right ] + \beta x + \gamma m +\delta \right )}, \,\, \text{if station $i$ does not detect the phase}\\
\label{eq:logEq}
\end{cases}
\end{align}

The coefficients $\alpha, \beta, \gamma, \delta$ in Equation \ref{eq:logEq} correspond to the regression coefficients that fit the data. $Dist\left [  \mathcal{L}, \mathcal{S}_i \right ] $ is the distance in degrees from $\mathcal{L}$ to $\mathcal{S}_i $, $x$ is the depth, and $m$ is the magnitude. Since we only consider one phase, we remove the phase index in $\mathbb{D}$ hereon. With this choice, we find the distance coefficient, $\alpha = -2.82$, the depth coefficient, $\beta = -0.03$, the magnitude coefficient, $\gamma = 1.14$, and the intercept, $\delta = 1.95$. From this we see that the distance and magnitude have a much higher influence than depth on the detection probability of the first P arrival.

\subsection{Arrival Time Model}
\subsubsection{Earth Model Uncertainty}\label{sec:modelnoise}
\label{sec:ttu}
    
\begin{figure}
\centering
\includegraphics[scale=0.24]{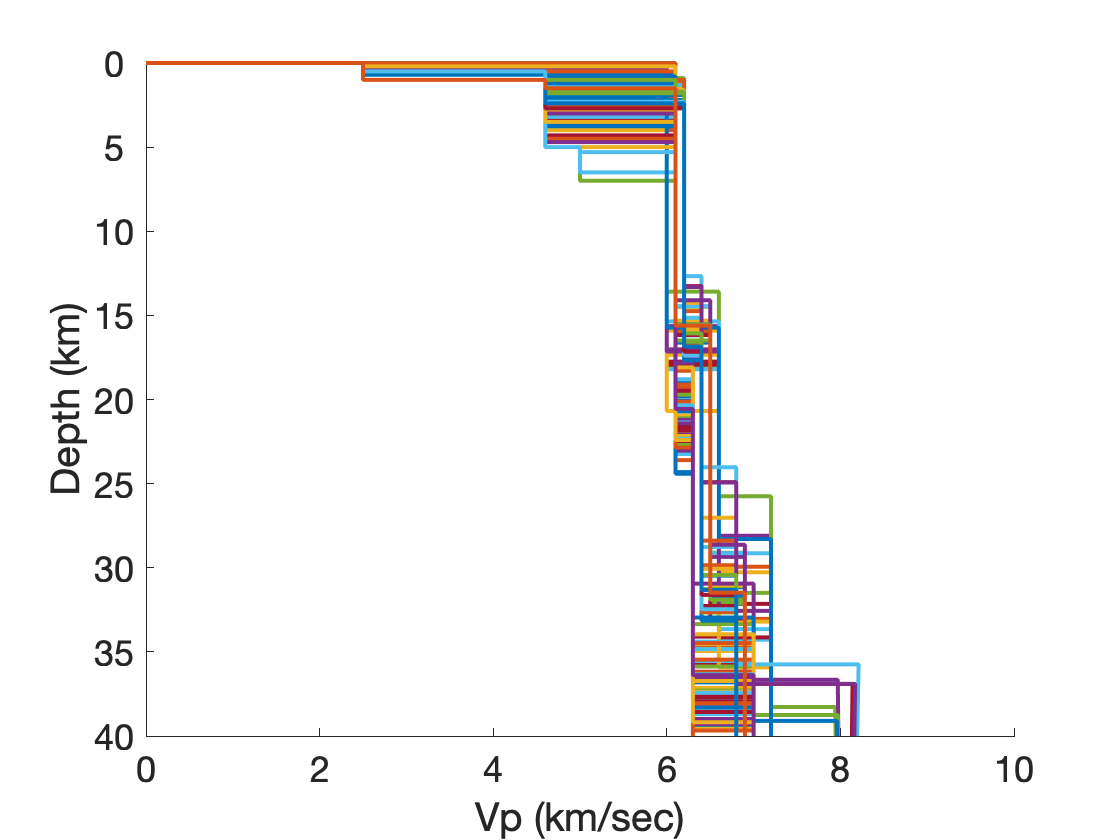}
\caption{Illustration of the 121 1D velocity models for Vp sampled from around the monitoring region in Figure \ref{fig:ta_map}. These representative earth models are used to estimate travel time uncertainty from earth model uncertainty.}
\label{fig:1d_models}
\end{figure}

For our travel time uncertainty model, first we will build an uncertainty model that captures the uncertainty due to the earth model.  We then will include a conditionally independent and additive measurement uncertainty model. We take the approach of using model uncertainty over using replicate variability because we are using synthetic earth models that produce the same output (up to measurement error) for the same inputs, although these earth models will have unknown errors when compared to potentially observable ``ground truth'' travel times. We treat this latent discrepancy between these models and the true experiment as aleatoric uncertainty since, in practice, experiments treat each event independently. It is possible to learn this discrepancy by jointly inferring events  \citep{myers2007bayesian}, however the added complexity is beyond the scope of this optimal experimental design study. For more details on this type of understanding, see \cite{kennedy2001computermodels} or \cite{maupin2020discrepancy}.

To capture earth model uncertainty, we selected 121 vertical cross-sections from Crust 1.0 \citep{laske2013update} from the area around the monitoring region to get 121 different 1D earth models with different Vp velocity profiles. These models can be seen in Figure \ref{fig:1d_models}. For each of these models we used TauP \citep{crotwell1999taup} to compute the travel times for different distances, $\Delta$, and depth, $x$, pairs. For a given distance and depth pair we compute the mean and variance of the travel times, $t_i$, given the travel times computed by TauP for the $N=121$ models:

\begin{align}
\mu \left ( \Delta, x \right ) &= \frac{1}{N} \sum_{i=1}^N t_i \left ( \Delta, x \right ) \label{eq:taupmodelmu} \\
\sigma \left (\Delta, x \right ) &= \sqrt{ \frac{1}{N-1} \sum_{i=1}^N [t_i \left ( \Delta, x \right ) -\mu \left ( \Delta, x \right )  ]^2}
\label{eq:taupmodelsig}
\end{align}

\begin{figure*}
\centering
\centerline{\includegraphics[scale=0.65]{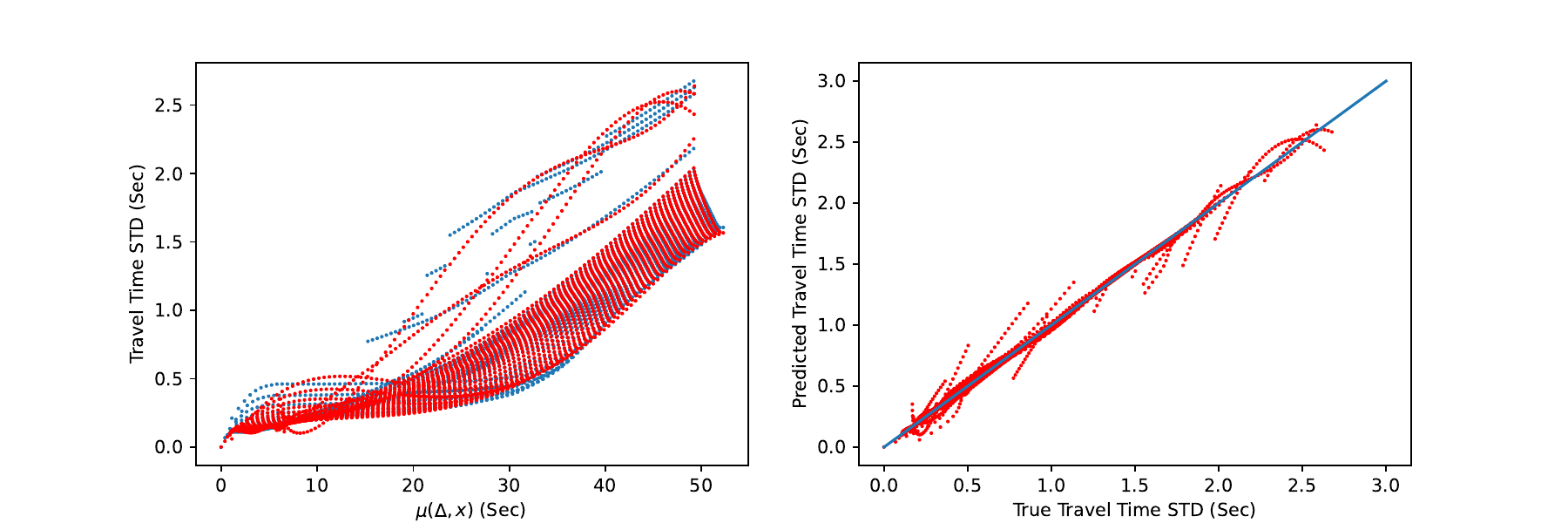}}
\caption{ \emph{Left}: A scatter plot (blue) of the estimated travel time mean $\mu \left ( \Delta, x \right )$, and estimated travel time standard deviation, $\sigma \left (\Delta, x \right )$ for various distance and depth pairs, superimposed with a fit polynomial model (red). \emph{Right}: A scatter plot of predicted travel time standard deviations compared to the true value (red). A line demonstrating the performance of a perfect model is displayed in blue. The polynomial model adequately captures the bulk trend, despite some variability due to the nature of the first arriving phase.}
\label{fig:tt_std}
\end{figure*}

Given the estimated mean and standard deviation pairs (Figure \ref{fig:tt_std}), we derive a model for the standard deviation of the travel time $\sigma_{model}$ as a $5$\textsuperscript{th} degree polynomial function of depth and distance that will be used in the likelihood.
This high-order polynomial sufficiently captures the major dependencies in the travel time standard deviation over the domain of interest but would fail to extrapolate beyond that domain. Therefore, care should be taken whenever using these types of function approximations that they are trained on the domain of interest as they are not intended for extrapolation. 
%$\sigma_{p} \left ( \Delta, x \right) = a\mu_p + b \mu_p^2 + c\mu_p^3$. 

\subsubsection{Measurement Error}\label{sec:measnoise}
As described earlier, we also develop a model of measurement uncertainty for each station that we treat as conditionally independent of the other station. This takes the form of phase pick uncertainty \citep{velascophasepick}, $\sigma_{meas}$, 
that depends on a sensor's signal-to-noise ratio (SNR). We model the SNR of a sensor's detection using a linear model that is a function of log distance, $\log\Delta$, and magnitude $m$, given by 
% we can use snr offset to tune sensor fidelity
% define SNR fit
\begin{equation}
    SNR = a\cdot m - b\cdot \log \Delta + c + \varepsilon,
\end{equation}
where
\[
\varepsilon \sim N(0,\sigma^2_{meas}{(SNR)}).
\]
We fit the coefficients $a,b$ and $c$ again using the Transportable Array dataset \citep{usarray}. We note that for our fit we found that no depth term was required which is why it was omitted but this will obviously depend on the problem context. We also add an offset term to this equation, potentially unique to each sensor, which allows us to tune sensor fidelity as we perform various experiments. As in \citep{velascophasepick}, the $\sigma_{meas}$ is thus given by

\begin{equation}
    \sigma_{meas}(SNR) =
        \begin{cases}
            \sigma_0 & \text{SNR} < t_L \\
            \gamma\sigma_0 & \text{SNR} > t_U \\
            \sigma_0 - \frac{\sigma_0 - \gamma\sigma_0}{\log(t_U) - \log(t_L)}\log(\frac{SNR}{t_L}) & \text{otherwise}
        \end{cases}
\end{equation}

\noindent where $\gamma$, $t_U$ and $t_L$, and $\sigma_0$ may all be tuned as hyperparameters (with $\gamma$ constrained to be less than 1). 

Ultimately, combining the modeling and measurement error we get that the total arrival error is
\[
  \sigma^2_{p}(\Delta, x, m) = \sigma^2_{model}(\Delta, x) + \sigma^2_{meas}(\Delta, m).
\].
\subsection{Travel Time Correlation}
    
While the previous models described the magnitude of errors at a station, they have not captured any correlations between the stations. In principle, we expect that there could be signification correlation in travel time uncertainty, particularly due to the earth model. We can compute the correlation between the travel times observed at two different stations at locations $\Delta_j$ and $\Delta_k$ for an event at depth $x$. This correlation is induced by the earth model uncertainty as:

\begin{equation}
\rho \left (\Delta_j,  \Delta_k, x \right ) = \frac{\sum_{i=1}^N [t_i \left ( \Delta_j, x \right ) -\mu \left ( \Delta_j, x \right )  ] [t_i \left ( \Delta_k, x \right ) -\mu \left ( \Delta_k, x \right )  ]}{\left (N-1 \right ) \sigma \left (\Delta_j, x \right ) \sigma \left (\Delta_k, x \right )}
\label{eq:corrcomp}
\end{equation}

\noindent Here $\mu$ and $\sigma$ are computed from the $N$ earth models in different locations from Crust 1.0 as in Equations \ref{eq:taupmodelmu} and \ref{eq:taupmodelsig}. For simplicity we will remove the depth dependence of the correlation by averaging the correlation over all $L$ depths. Therefore we estimate the correlation between two sensors as $\rho \left (\Delta_j,  \Delta_k \right )  = \frac{1}{L} \sum_{l=1}^L \rho \left (\Delta_j,  \Delta_k, x_l \right )$.

\begin{figure*}
\centering
\includegraphics[scale=0.25]{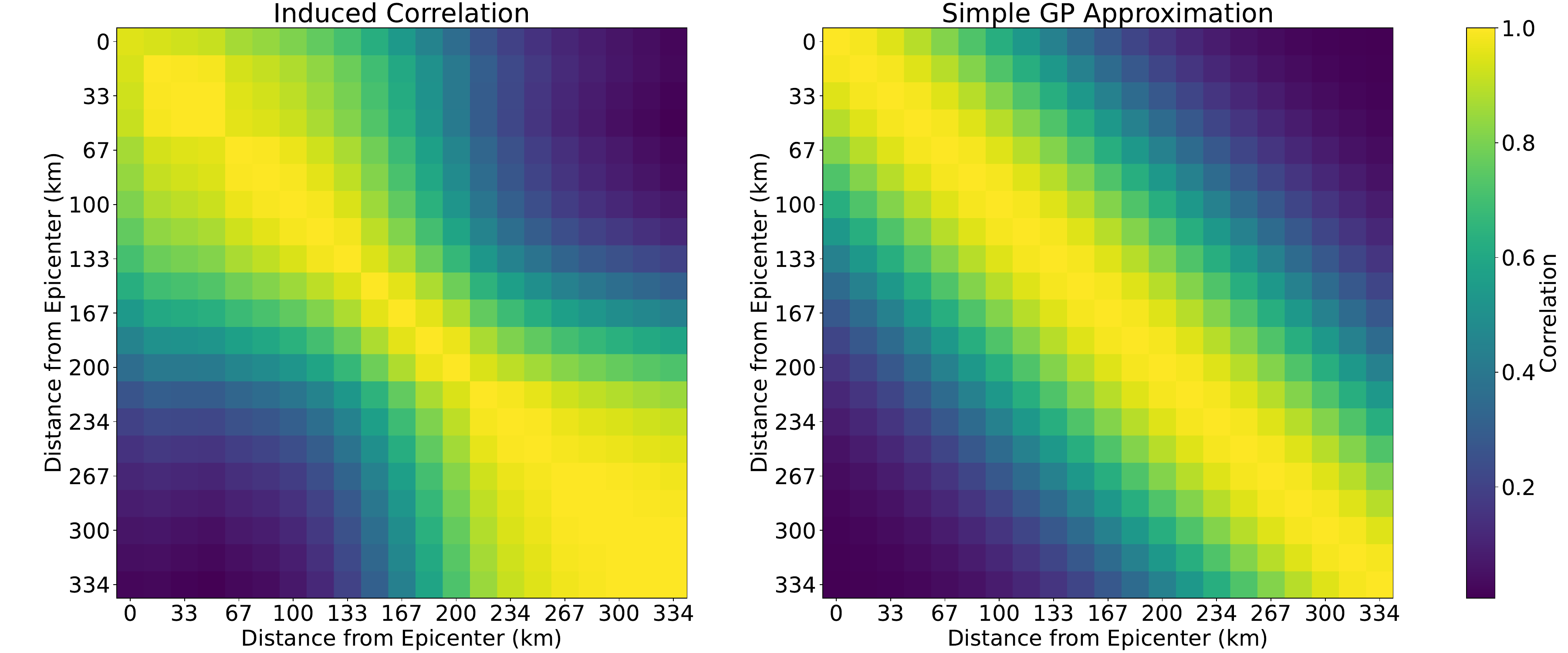}
\caption{Visualization of the correlation matrix. The left figure is the correlation matrix, $\Gamma$, induced by the earth model uncertainty computed using Equation \ref{eq:corrcomp}. The axes correspond to the distance along the surface from the source epicenter in km. Stations are approximately spaced 33km apart. The right figure is a simplified model of the correlation, $\Gamma_{GP}$, found using a square- exponential kernel, which approximates $\Gamma$. The square-exponential kernel assumes that correlation is only a function of the distance between stations. This simplified model reasonably captures the correlation length scale for $\Gamma$ but is unable to capture the complex, non-translation invariant block structure of $\Gamma$. }
\label{fig:gp_corr}
\end{figure*}

We define the full correlation matrix, $\Gamma$, between the stations at distances $\Delta_i$ from the source has having elements $\Gamma_{jk} = \rho \left (\Delta_j,  \Delta_k \right )$. We want to fit a Gaussian process model with a square exponential kernel to this data so we can easily estimate the correlation between arbitrary sensor pairs when designing the network i.e. we want $\Gamma \approx \Gamma_{GP}$. Therefore we want to find the a correlation length, $\mathfrak{l}$, such that $\left \{ \Gamma_{GP} \right \}_{jk} = \exp \left [ -\frac{1}{2 \mathfrak{l}^2} \left (\Delta_j - \Delta_k \right )^2 \right ]$ and $\Gamma_{GP}$ minimizes the discrepancy with $\Gamma$. We, under our modeling conditions, find the correlation length scale as $\mathfrak{l} = 147.5$ km. The comparison of $\Gamma$ and the resulting $\Gamma_{GP}$ can be seen in Figure \ref{fig:gp_corr}. We observe that the square exponential kernel is able to capture the general length scale of the induced correlation, meaning that stations that are close together are more correlated, but does not capture its complexity. The induced correlation has a block-like structure where stations that are near to the source are highly correlated, stations far from the source are highly correlated, and stations in the transition region exhibit less strong correlation with nearby stations. This likely corresponds to the type of first arrival that is being observed at each station, where close stations observe a Pg while far stations observe a Pn. Our choice of GP kernel is translation invariant meaning that the sensor correlation is only a function of the distance between the two sensors and does not depend on the source parameters. More generally, a different GP kernel would need to be constructed for each seismic source, which is computationally challenging for the nested complexity of OED. Considering only a translation invariant GP kernel is obviously a simplification but provides a first step towards modeling station correlation which is typically very difficult and often ignored.

We can now construct the P arrival time likelihood $p \left(  \mathbb{A} \mid  \mathcal{L}, x, t_o, \mathbb{D}, \mathcal{S}  \right )$ by combining our model of the travel time prediction, $\mu_p$; earth-model-induced standard deviation, $\sigma_{model}$; measurement-induced standard deviation $\sigma_{meas}$; and correlation matrix, $\Gamma_{GP}$:

\begin{align}
p &\left(  \mathbb{A} \mid  \mathcal{L}, x, t_o, \mathbb{D}, \mathcal{S}  \right )= \nonumber\\
&\frac{1}{\left (2 \pi \right )^{|\mathbb{D}|/2} | \Sigma |^{1/2}} \exp \left ( - \frac{1}{2} \left [\mathbb{A} - \mu_p \left ( \mathcal{L}, x, \mathbb{D}, \mathcal{S} \right )- t_o  \right ]^T \Sigma^{-1} \left [\mathbb{A} - \mu_p \left ( \mathcal{L}, x, \mathbb{D}, \mathcal{S} \right )- t_o  \right ] \right ) \label{eq:finalgpmodel} \\
\Sigma &= \sigma_{model}^T\left(\Delta, x, \mathbb{D} \right)\Gamma_{GP}(\mathbb{D}, \mathcal{S})\sigma_{model}\left(\Delta, x, \mathbb{D} \right) + \text{diag}\left[\sigma^2_{meas}(\Delta, m, \mathbb{D}, \mathcal{S})\right]
\end{align}

\noindent Here $|\mathbb{D}|$ is the number of detections, $| \Sigma |$ is the determinant, $\mu_p \left ( \mathcal{L}, x, \mathbb{D}, \mathcal{S} \right )$ is a vector of the predicted travel times for stations that had a detection, $\sigma_{model} \left (\Delta, x, \mathbb{D} \right )$ is a vector of the predicted standard deviations of the travel time to each station induced by the earth model uncertainty, and $\text{diag}\left[\sigma^2_{meas} \left (\Delta, m, \mathbb{D}, \mathcal{S} \right )\right]$ is a diagonal matrix of the squared predicted standard deviations of the travel time to each station induced by the measurement uncertainty. Finally, $\Gamma_{GP} \left ( \mathbb{D}, \mathcal{S} \right )$ is the estimated correlation between stations using the GP model.

We further note that we marginalized our source origin time prior over $t_o$, assuming a uniform improper prior (meaning that an event is equally likely at any time), and therefore omit it from the model. We assume this prior since the origin time is naturally restricted by the size of the chosen domain. The improper uniform prior only behaves differently from a proper prior on the edges of the proper prior's domain, so any uniform prior that is wide enough to ensure that possible travel times given the domain do not occur on the edges of the domain should be functionally equivalent to an improper prior. See Figure \ref{fig:improper_prior} for further details. This reduces the dimension of our seismic source parametrization space and leads to our final model of the arrival time likelihood:

\begin{align}
\label{eq:finalTlike}
p &\left(  \mathbb{A} \mid \mathcal{L}, x, \mathbb{D}, \mathcal{S}  \right ) = \int_{t_o} p \left(  \mathbb{A} \mid \mathcal{L}, x, t_o, \mathbb{D}, \mathcal{S}  \right ) p \left(  t_o \right) dt_o \nonumber\\
&= \frac{1}{\left (2 \pi \right )^{(|\mathbb{D}|-1)/2} | \Sigma |^{1/2} \beta^{1/2}} \exp \left (- \frac{1}{2} \left [ \mathbb{A} - \mu_p \left ( \mathcal{L}, x, \mathbb{D}, \mathcal{S} \right ) \right ]^T \Sigma^{-1} \left [ \mathbb{A} - \mu_p \left ( \mathcal{L}, x, \mathbb{D}, \mathcal{S} \right ) \right ]\right ) \exp \left (\frac{\alpha^2}{2\beta} \right )
\end{align}

\begin{align}
\alpha &= \mathbbm{1}^T \Sigma^{-1}  \left [ \mathbb{A} - \mu_p \left ( \mathcal{L}, x, \mathbb{D}, \mathcal{S} \right ) \right ] \\
\beta &=  \mathbbm{1}^T \Sigma^{-1}  \mathbbm{1} 
\end{align}

\section{Computational Approach}
\label{sec:comp}
\subsection{Estimating Information Gain}

Given the Bayesian framework introduced in Section \ref{sec:Bayes} and the specific models introduced in Section \ref{sec:BSM} we present a method to estimate the expected information gain, $\mathcal{I} \left ( \mathcal{S} \right )$, of the sensor network $\mathcal{S}$. Recall from Equation \eqref{eq:expectKL} that we can express EIG as:

\begin{equation}
\mathcal{I} \left ( \mathcal{S} \right ) = \int p \left ( \theta^\prime \right ) \int p \left ( \mathcal{D} \mid \theta^\prime, \mathcal{S} \right) \int p \left (\theta\mid \mathcal{D},   \mathcal{S}  \right ) \log \frac{p \left (\theta\mid \mathcal{D},  \mathcal{S}  \right )}{p \left (\theta\right )} d\theta d\mathcal{D} d\theta^\prime
\end{equation}

\noindent We can further define $\mathcal{I} \left ( \mathcal{S} \mid \theta^\prime \right )$ as the expected information gained about a specific event $\theta^\prime$ where
\begin{equation}
\label{eq:thetaEIG}
\mathcal{I} \left ( \mathcal{S} \mid \theta^\prime \right ) =\int p \left ( \mathcal{D} \mid \theta^\prime, \mathcal{S} \right) \int p \left (\theta\mid \mathcal{D},   \mathcal{S}  \right ) \log \frac{p \left (\theta\mid \mathcal{D},  \mathcal{S}  \right )}{p \left (\theta\right )} d\theta d\mathcal{D}
\end{equation}
and thus express $\mathcal{I} \left ( \mathcal{S} \right )$  as:
\begin{equation}
\label{eq:fullEIG}
\mathcal{I} \left ( \mathcal{S} \right ) = \int  \mathcal{I} \left ( \mathcal{S} \mid \theta^\prime \right )  p \left ( \theta^\prime \right ) d\theta^\prime
\end{equation}

\noindent $\mathcal{I} \left ( \mathcal{S} \mid \theta^\prime \right )$ is an important quantity on its own as it can be used to tell how sensitive the network is to a specific event $\theta^\prime$ We can then produce maps of this sensitivity in order to communicate how the network performs under different conditions in order to check against requirements.

We will use the approach of estimating  $\mathcal{I} \left ( \mathcal{S} \mid \theta^\prime \right )$ to estimate EIG. First, we draw samples from $\theta'$ from $p(\theta')$ to construct a large set of candidate seismic events using a method like importance sampling with a Quasi Monte Carlo (QMC) mesh \citep{ref:mcbook}. QMC provides an efficient set of space-filling samples that requires significantly fewer samples than a standard uniform grid. Then, for each element in our event space, we estimate $\mathcal{I} \left ( \mathcal{S} \mid \theta^\prime \right )$ and average them to estimate $\mathcal{I} \left ( \mathcal{S} \right )$.  To estimate  $\mathcal{I} \left ( \mathcal{S} \mid \theta^\prime \right )$ we construct hypothetical datasets by sampling $p \left ( \mathcal{D} \mid \theta^\prime, \mathcal{S} \right)$. Then we will solve the Bayesian inference problem given the datasets to estimate the information gain measured via the KL divergence. We solve the Bayesian inference problem over the discrete event space instead of a continuous event space for computational efficiency, although this results in a bias. Solving the Bayesian inference problem involves sampling from the prior distribution, which we accomplish using importance sampling (discussed further in section \ref{sec:importance_sampling}). As long as enough discrete points are used, the KL divergence will converge to the same value as the continuous distribution so the bias will be small. By looking at statistics of the posterior probabilities of the discrete events, we can assess whether enough points have been used. The hypothetical data are constructed by sampling $p \left(  \mathbb{A} \mid \mathcal{L}, x, \mathbb{D}, \mathcal{S}  \right )$ and $p \left( \mathbb{D} \mid \mathcal{L}, x, m, \mathcal{S}  \right ) $. The resulting algorithm is summarized in Algorithm 1.
\begin{algorithm}
\caption{Expected Information Gain (EIG) Calculation}
\begin{algorithmic}[1]
\State \textbf{Input:} $\mathcal{S}$ (sensor configuration), $\Theta$ (plausible events), $p(\theta')$
\State \textbf{Result:} $I(\mathcal{S} | \theta)$ for individual events $\theta$, and total $I(\mathcal{S})$
\For{each event hypothesis $\theta' \in \Theta$}
    \State simulate arrival dataset according to $D \sim p(D | L', x', m', \mathcal{S})$
    \For{each arrival dataset $D$}
        \State simulate arrival time according to $A \sim p(A | L', x', D, \mathcal{S})$
        \For{each simulated dataset $D = \{A, D\}$}
            \State compute likelihood $p(D | \theta, \mathcal{S})$ using Equations 9 and 17
            \State compute posterior $p(\theta | D, \mathcal{S}) \propto p(D | \theta, \mathcal{S}) p(\theta)$
            \State compute KL divergence for information gain $I(\mathcal{S} | \theta', D)$
        \EndFor
    \EndFor
    \State compute EIG for $\theta'$, as average of $I(\mathcal{S} | \theta', D)$ across simulated data
\EndFor
\State compute total EIG $I(\mathcal{S})$ as average over all event hypotheses and data
\end{algorithmic}
\end{algorithm}

We choose to use this approach as opposed to a Markov chain Monte Carlo (MCMC) method for two reasons. First, the dimension of the sample space is small, allowing us to draw enough samples to reliably reconstruct the prior and posterior distributions. Second, this approach allows us to reuse likelihood computations for each sample across all steps of the algorithm, whereas an MCMC method would require computing a new likelihood at each iteration. In applications where the dimension of the sample space is higher, an MCMC method would likely be preferred. We also acknowledge that there are potential issues with this approach in cases where the importance distribution does a poor job of approximating the sampling distribution \citep{williams2021sensoraug}. This could be particularly problematic in cases where diffuse prior samples are used for sampling a concentrated posterior.
% sampling when posterior samples are required as in Equations \ref{eq:expectKL} and \ref{eq:thetaEIG}.
In further work we hope to explore alternative sampling methods such as MCMC, double-nested Monte Carlo, and those discussed in \cite{picard2019postdensity} and compare their performance to the method used in this work.

\subsection{Optimization}
    
Once we have the algorithms to estimate $\mathcal{I} \left ( \theta^\prime \mid \mathcal{S} \right )$ and $\mathcal{I} \left (  \mathcal{S} \right )$, we can formulate the optimal experimental design problem to choose the the location and type of different seismic stations. We can use the greedy Bayesian optimization method described in \ref{sec:bayesopt}. We use the Python library \textsc{scikit-opt} \citep{head_tim_2020_4014775}
to implement Bayesian optimization with a Gaussian Process (GP) surrogate. This library enables us to adaptively learn hyperparameters of the GP kernel function e.g. length scales of the squared exponential kernel, magnitudes of the additive white noise, etc. Further, it can support several different criteria for Bayesian optimization that control the way in which the optimizer balances exploration versus exploitation in Bayesian optimization. This tradeoff means that the Bayesian optimizer has to choose sample points that enable it to both learn the surrogate for the EIG surface and find points that optimize the EIG. The common criteria for this found in \textsc{scikit-opt} are the Expected Improvement, Lower Confidence Bound, and Probability of Improvement. \textsc{scikit-opt} also has the option to mix these criteria and choose one at random. We found that Expected Improvement works well but have not systematically explored all these options.

\subsection{Software Implementation}

The models in Section \ref{sec:BSM} and algorithms from this section can be found on GitHub \citep{catanach2024github}. This code provides the tools necessary to analyse and optimize seismic monitoring networks. Currently we target the location problem, like those discussed in Section \ref{sec:results}, in which we want to study how well the network will identify the location of an event and then optimize the network to provide better locations. A detailed explanation of the software implementation can be found in \ref{sec:appdx_software}. 

\section{Results}
\label{sec:results}
Unless otherwise specified, we explore a simple model for placing sensors to monitor a square domain for latitudes between
40\textdegree N and 42\textdegree N, longitudes between 112\textdegree W and 108:36\textdegree W, magnitudes greater than 0. and depth between 0km and 40km. For computing the EIG, 10k events were chosen using a Quasi Monte Carlo
(QMC) mesh defined by the Sobol sequence. For each candidate event 32 hypothetical data
realizations were used. 100 steps of Bayesian optimization were used to optimize the the sensor
configuration.

\subsection{Prior distributions}
We perform our experiments with one or both of the following prior distributions on our seismic parameters.

The first prior used was a uniform prior. Under this prior, seismic sources are assumed to have a uniform prior probability in this domain. We also assume
    that the magnitude prior is an exponential distribution with rate parameter $\lambda = \log(10)$ and a
    minimum magnitude of 0.5. This prior means that the likelihood of an event of a given magnitude
    falls off exponentially as the magnitude increases. We assume that the origin time is a uniform
    improper prior meaning that all times are equally likely. The sensors are also limited to be placed
    in this domain. 
    
The second distribution used a mixture distribution on latitude and longitude to very simply simulate both a fault line and a point source. It used a uniform distribution on depth and an exponential distribution with $\lambda=10$ on magnitude. 

\begin{figure}
    \centering
        \includegraphics[width=.6\linewidth]{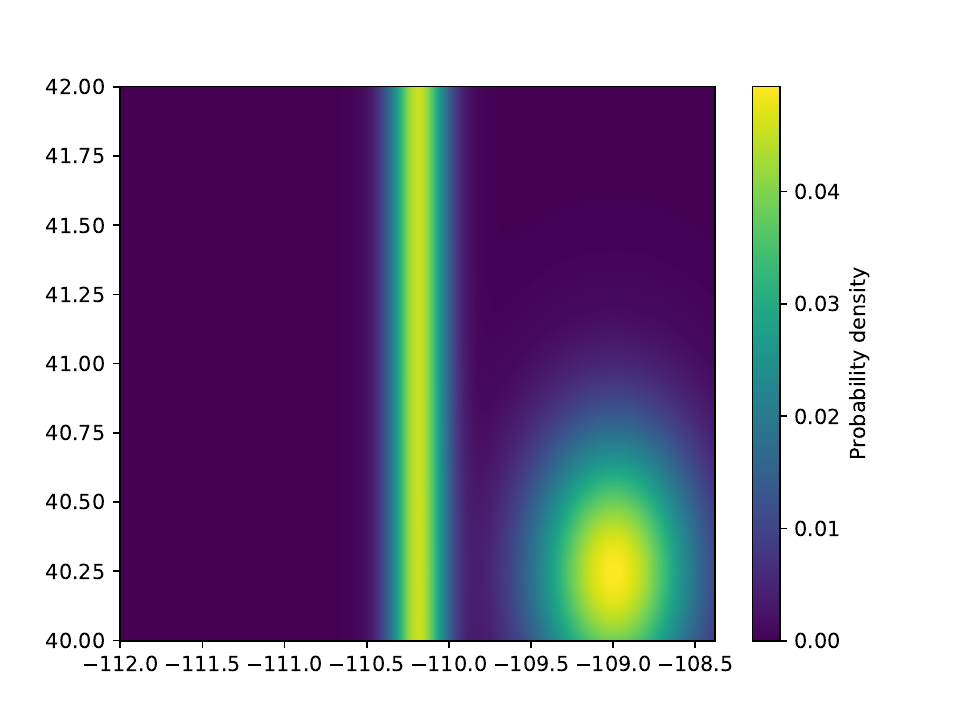} \caption{Simple prior distribution on latitude and longitude representing a fault and point source. It is comprised of 3 mixture components: a bivariate Gaussian representing the point source, a univariate Gaussian in the longitude direction multiplied by a uniform in the latitude direction representing the fault, and a uniform in both directions representing the background probabilities. }\label{fig:faultbox_prior}
    \end{figure}

    We choose the mixture distribution on latitude and longitude to represent a fault line and a point source. The first mixture component is a bivariate Gaussian centered at (40.25,-109) with covariance matrix 
    \[\Sigma = \left[\begin{matrix} .125 & 0 \\
                        0 & .125
    \end{matrix}\right].\]
    The second mixture component is a 1-dimensional Gaussian in the longitude direction with mean -110.19 and standard deviation .125 multiplied by a uniform in the latitude direction. The final mixture component is a uniform distribution across both latitude and longitude. These components were given mixture weights .49, .49, and .02 respectively. See Figure \ref{fig:faultbox_prior} for a visual representation. 

    The total probability for a single event under this prior is thus given by the product of the probability for each parameter. For convenience, we refer to this second prior as the fault-box prior.

\subsection{Analysis results}
\begin{figure}
    \centering
    \includegraphics[width=.7\textwidth]{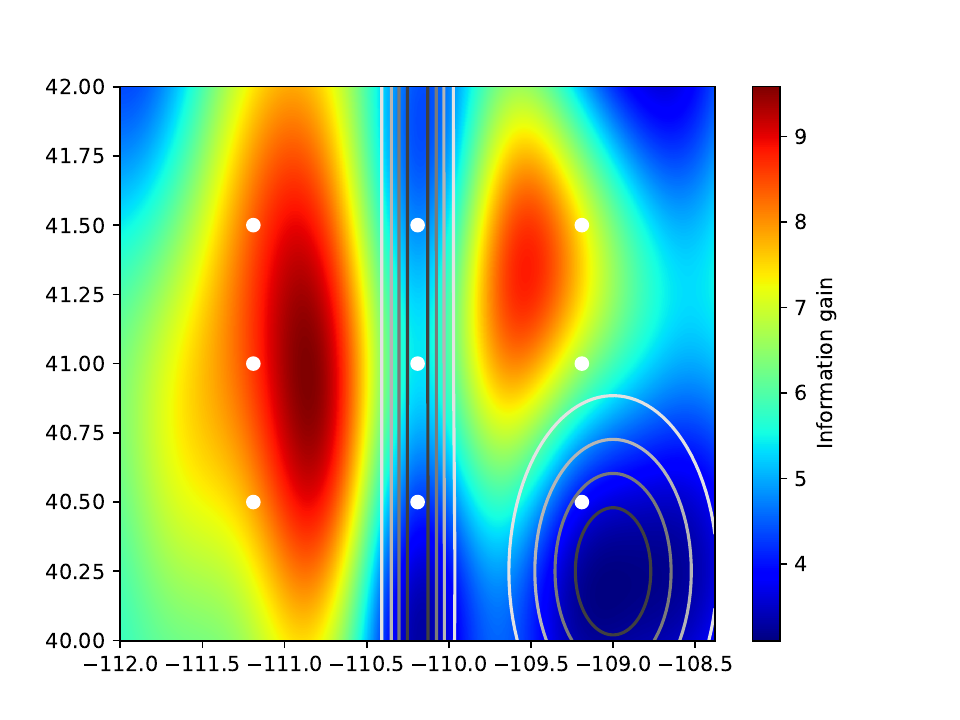}
    \caption{A network sensitivity analysis to events in a given latitude/longitude domain. The prior distribution on event location is shown by the gray contour plot. Events that are far from the high-density areas of the prior distribution contribute more information than events close to high-density areas.}
    \label{fig:single_heatmap}
\end{figure}

Using our algorithm, we can perform two different types of analyses: We can design new sensor networks for a given area, and we can analyse the sensitivity of existing sensor networks to events in a given area. Figure \ref{fig:sequential_placement} shows what it looks like when sensors are placed sequentially in a given area. Figure \ref{fig:single_heatmap} shows an an analysis of network sensitivity to events in a given area. We see that the network gains more information about events that are far from the high-density areas of the prior, and less information about events that are closer to high-density areas. 

\subsection{Effect of Sensor Fidelity}
\begin{figure*}
\begin{subfigure}[b]{.5\textwidth}
\centering
\includegraphics[width=1\textwidth]{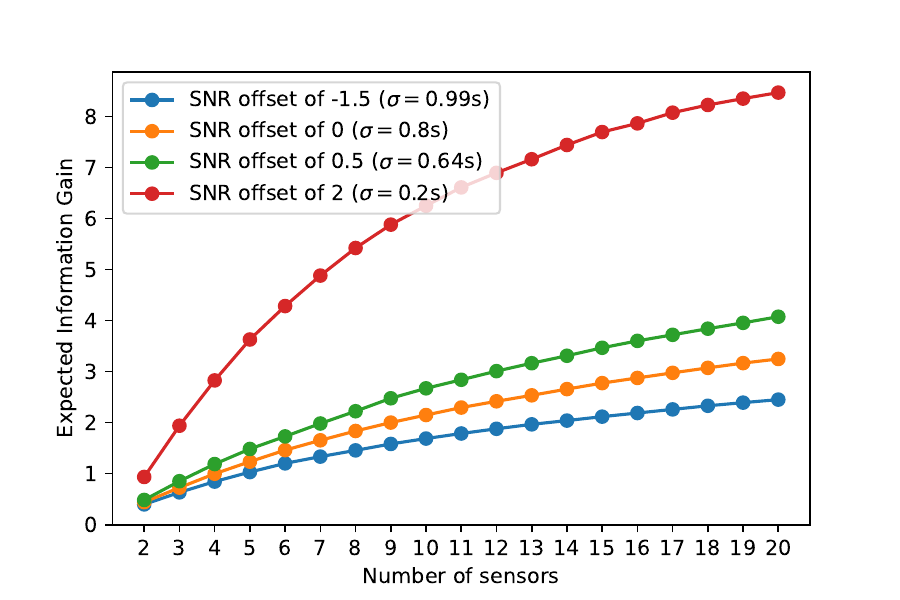}
\end{subfigure}%
\begin{subfigure}[b]{.5\textwidth}
\centering
\includegraphics[width=1\textwidth]{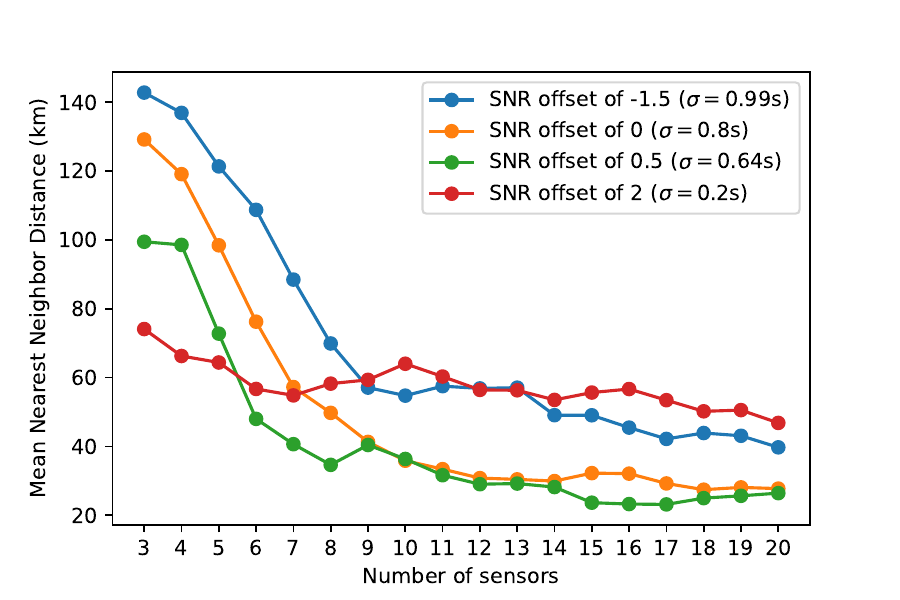}
\end{subfigure}%
\caption{The left panel illustrates the change in EIG as additional sensors are placed
    using greedy optimization for three different networks with different sensor fidelities.
    These networks have signal-to-noise (SNR) ratio offsets of 2 seconds, .5 seconds, 0 seconds, and -1.5 seconds. The corresponding average measurement error for these networks, $\sigma$, is listed in parentheses. See Table \ref{tab:std_conversions} for a description of the relationship between SNR and measurement error. The right
    panel describes the geometry of the networks based upon how close the stations are
    to each other. We can see that, particularly in the beginning, the noisier the network is, the farther apart stations are added to those networks.}
\label{fig:fid_curves}
\end{figure*}

We next investigate how sensor placement is affected by varying sensor fidelity conditions. We control the fidelity of a given sensor by adding an offset to its ratio of signal to measurement noise, an offset than can be thought of as corresponding to measurement noise with a given standard deviation. Using a uniform prior, we place 20 sensors using four different sensor fidelity values (See Figure \ref{fig:fid_curves}). Unsurprisingly, we see that as sensor fidelity increases, the network's information gain also increases. It is difficult to see a clear pattern in sensor proximity, but we notice that as sensor fidelity increases, sensors are generally placed closer together. This could be due to the fact that noisy sensors need to be placed farther apart from each other than less noisy sensors in order to properly triangulate events.

\begin{figure*}
% TODO: Switch axis and log scale on x
\begin{subfigure}[b]{.5\textwidth}
\centering
\includegraphics[width=.95\textwidth]{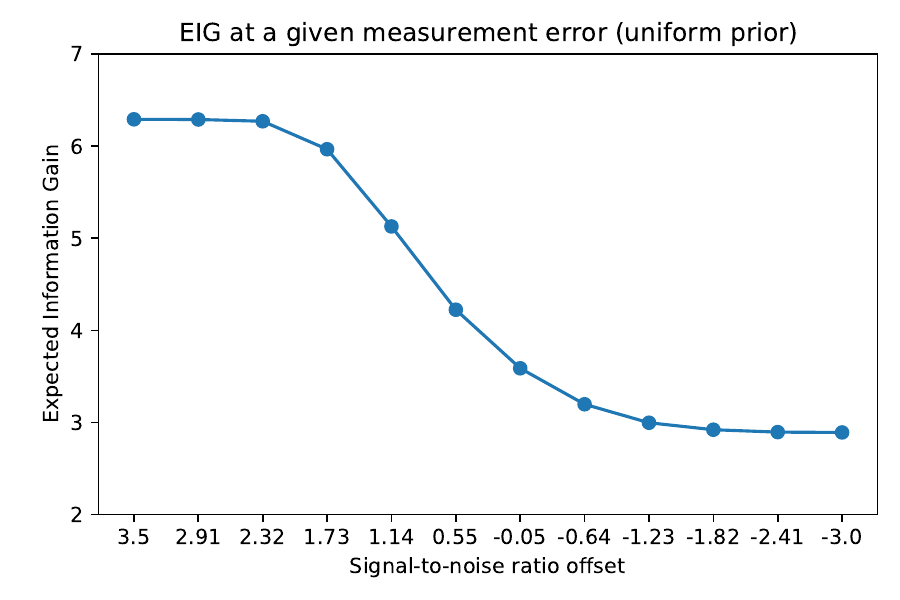}
\end{subfigure}%
\begin{subfigure}[b]{.5\textwidth}
\centering
\includegraphics[width=.95\textwidth]{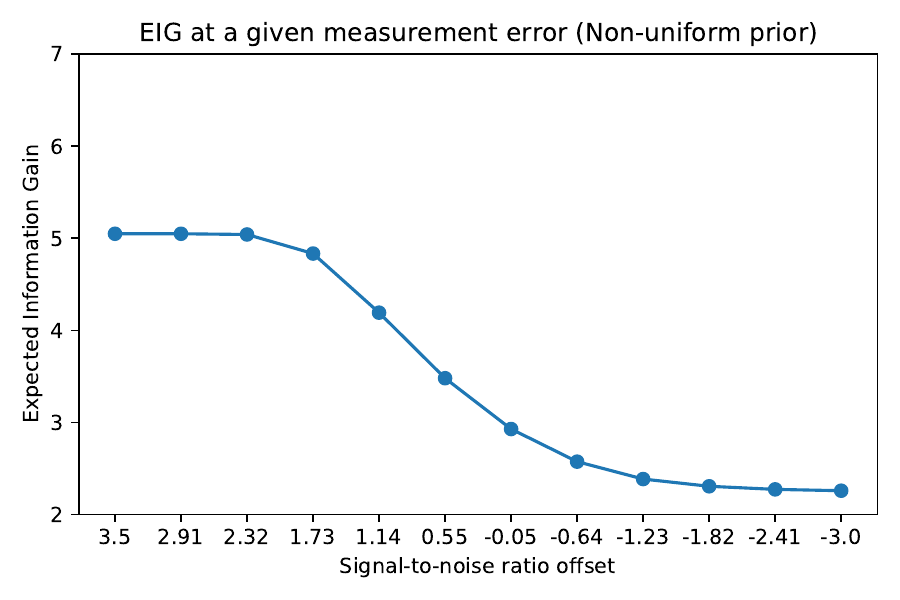}
\end{subfigure}%
\caption{Illustrations the degradation of EIG for all events on both priors as the signal-to-noise ratio is decreased. This analysis shows where measurement error dominates versus modeling error and vice versa. We see that EIG is fairly stable when SNR is offset by more than 1.73 or less than -0.64, and is greatly affected when the signal offset is between those values. See Table \ref{tab:std_conversions} for a description of the relationship between SNR and measurement error. For the average event, EIG is more sensitive to the measurement error when the SNR value is between 1.73 and -0.64. These larger fluctuations in measurement error correspond to the steeper curve in the plots. On the other hand, an SNR offset greater than 1.73 means that model error dominates, while an SNR offset of less than -0.64 means that measurement error dominates. This can inform where to invest effort in reducing uncertainty.}
\label{fig:eig_sidebyside}
\end{figure*}

We next investigate the effect of sensor fidelity on information gain. We examine the effect of fidelity by comparing the information gain surface generated by a grid of 9 evenly spaced sensors across 12 different signal-to-measurement noise ratio offsets. In this experiment, these evenly spaced offsets range from -3.0 to 3.5. We perform this experiment using both a uniform prior on events and the non-uniform fault-box prior. The results of these experiments can be seen in Figure \ref{fig:eig_sidebyside}, and full visualizations of how the SNR affects IG across all events in a domain can be found in Figures \ref{fig:fid_surface_uniform} and \ref{fig:fid_surface_faultbox}.  As when controlling the measurement noise standard deviation directly, we see that below a certain fidelity offset value the measurement noise dominates the signal and as such we see minimal information gain. Once past a certain threshold (in this case a sensor fidelity of -.045 corresponding to a measurement noise standard deviation of 1.59) the model uncertainty begins to take over and we see an increase in information gain in both the uniform and non-uniform prior cases.

\begin{table}
	    \centering
	   % \bigskip
	    \begin{tabular}{|c|c|c|}
	    \hline
		SNR offset &$\sigma_{meas}$, uniform prior & $\sigma_{meas}$, nonuniform prior \\ 		\hline \hline
		3.5  &  0.1   &  0.1\\% \hline
		2.91  & 0.1   &  0.1\\% \hline
		2.32  & 0.13   &  0.15\\ %\hline
		1.73  & 0.23   &  0.38\\% \hline
		1.14 & 0.41   &  0.75\\% \hline
		0.55 & 0.62 & 1.19\\% \hline
		-0.05 & 0.81 & 1.5\\% \hline
		-0.64 & 0.92 & 1.83\\% \hline
		-1.23 & 0.97 & 1.94\\% \hline
		-1.82 & 0.99 & 1.98\\% \hline
		-2.41 & 1.0 & 1.99\\ %\hline
		-3.0 & 1.0 & 2.0 \\ \hline
	    \end{tabular}
	    	    \caption{Table displaying the effect of SNR offset on measurement error for both the uniform and nonuniform prior. The first column shows the offset value, and the second two columns show the average measurement error across all events sampled from the given prior for sensors with the given offset. It appears that between values of 1.73 and -0.64, the average measurement error is more sensitive to changes in SNR. We emphasize that this is not an equivalency table, but rather a notional description of how the measurement error changes as the SNR is changed.}
	    \label{tab:std_conversions}
	\end{table}
	
\subsection{Optimizing a network with boundary constraints}\label{sec:boundary_opt_example}
\begin{figure*}
\begin{subfigure}[b]{.5\textwidth}
\centering
\includegraphics[width=.95\textwidth]{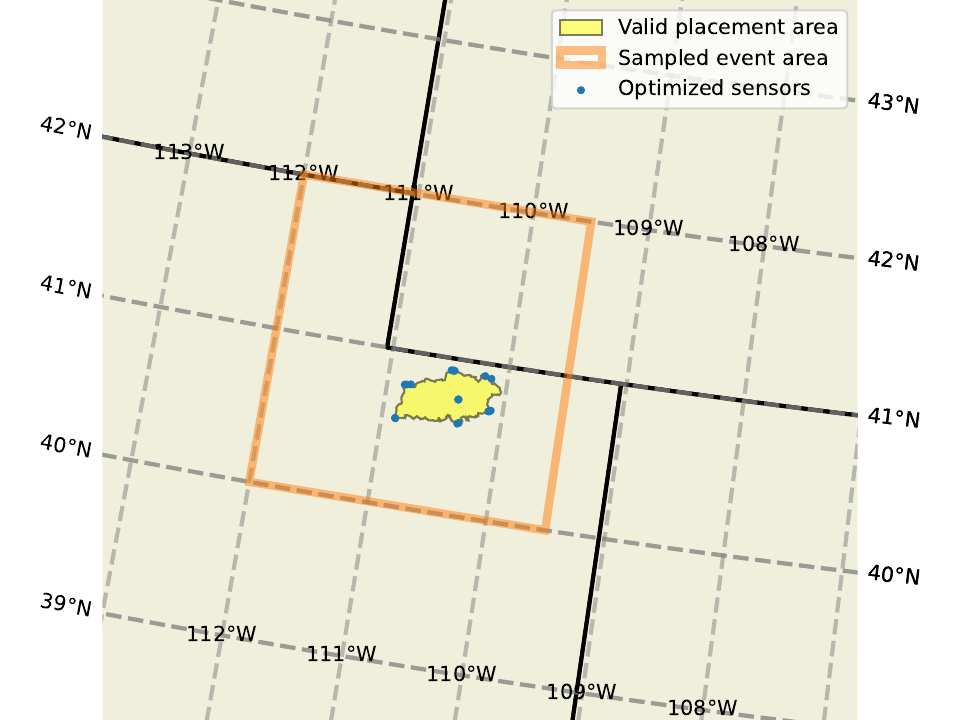}
\end{subfigure}%
\begin{subfigure}[b]{.5\textwidth}
\includegraphics[width=.95\textwidth]{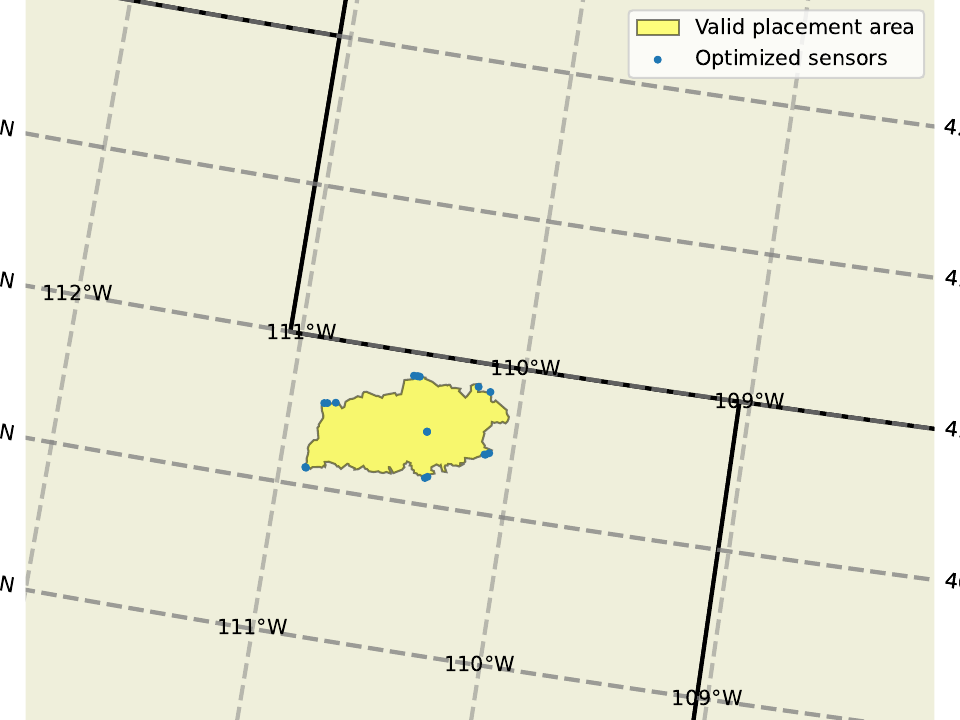}
\end{subfigure}%
\caption{Network optimized under the shown boundary constraints. Twenty sensors were placed within the highlighted yellow area. The figure on the left shows the domain on which the seismic models were trained (the pink area in Figure \ref{fig:ta_map}), with the orange highlighted box on the left being the area from which events were sampled (the orange area in Figure \ref{fig:ta_map}). The figure on the right shows just the area in which the events were sampled. We see that sensors were placed near the boundary of the admissible area. This is possibly done in an attempt to better capture events outside the optimization domain.}
\label{fig:uinta_opt}
\end{figure*}

        We also examined the behavior of the optimization when constraints were placed on the location of the sensors according to Figure \ref{fig:uinta_opt}. This boundary was chosen based on the boundaries of the Uinta National Forest, which was chosen because the Uinta National Forest are irregular and therefore provide a good test for the bounded optimization software, and they are also entirely contained within the area on which our models were trained.
        
        The network created by our script under our boundary constraints is shown in Figure \ref{fig:uinta_opt}. We can see that sensors are placed on the edges of boundaries in order to gain information about the surrounding area.

\subsection{Effect of Correlation}
\begin{figure*}
\begin{subfigure}[b]{.5\textwidth}
\centering
\includegraphics[width=.95\textwidth]{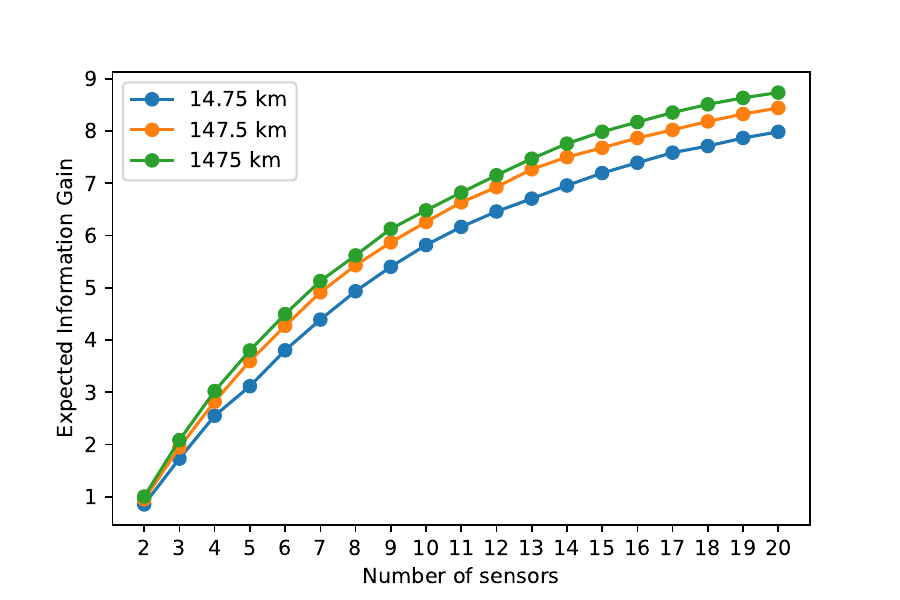}
\end{subfigure}%
\begin{subfigure}[b]{.5\textwidth}
\centering
\includegraphics[width=.95\textwidth]{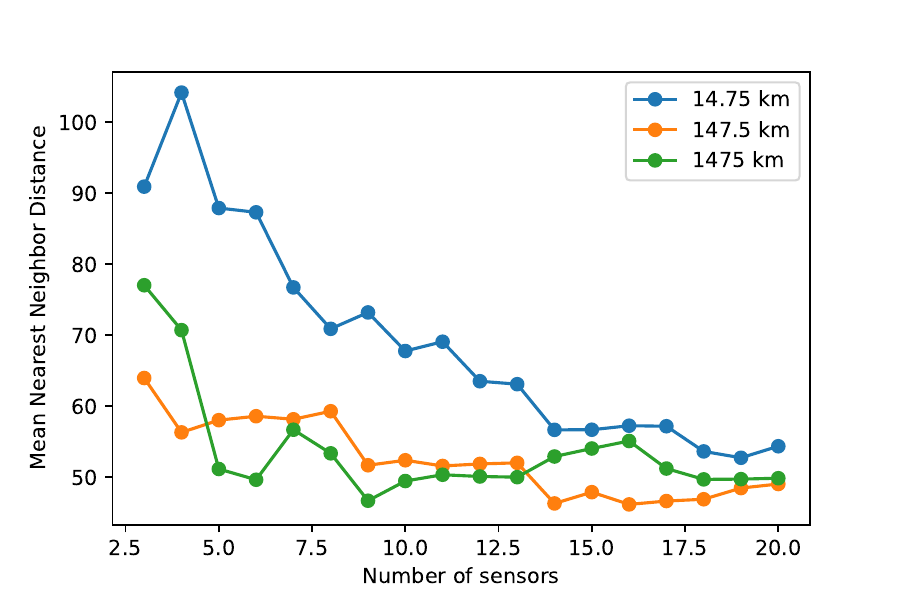}
\end{subfigure}%

\begin{subfigure}[b]{.5\textwidth}
\includegraphics[width=.95\textwidth]{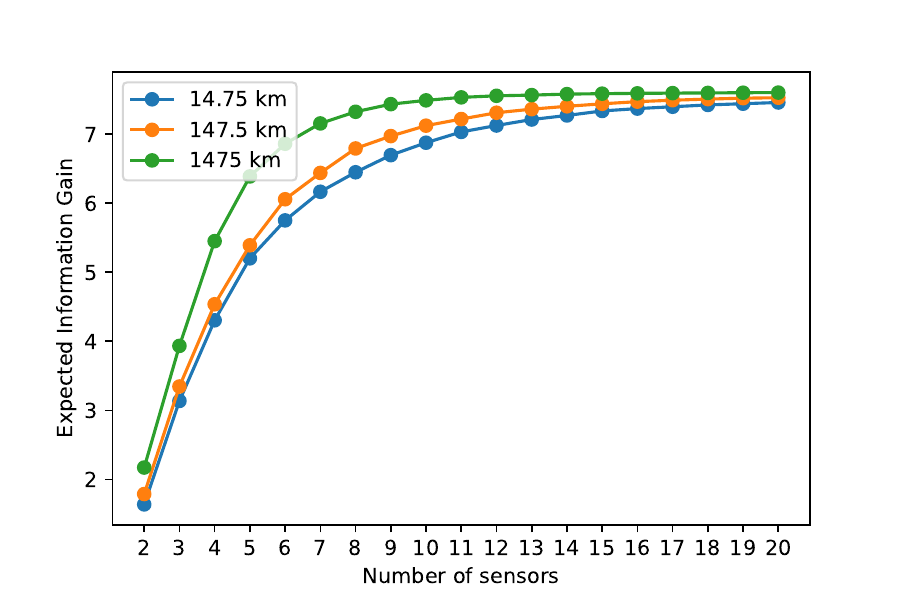}
\end{subfigure}%
\begin{subfigure}[b]{.5\textwidth}
\centering
\includegraphics[width=.95\textwidth]{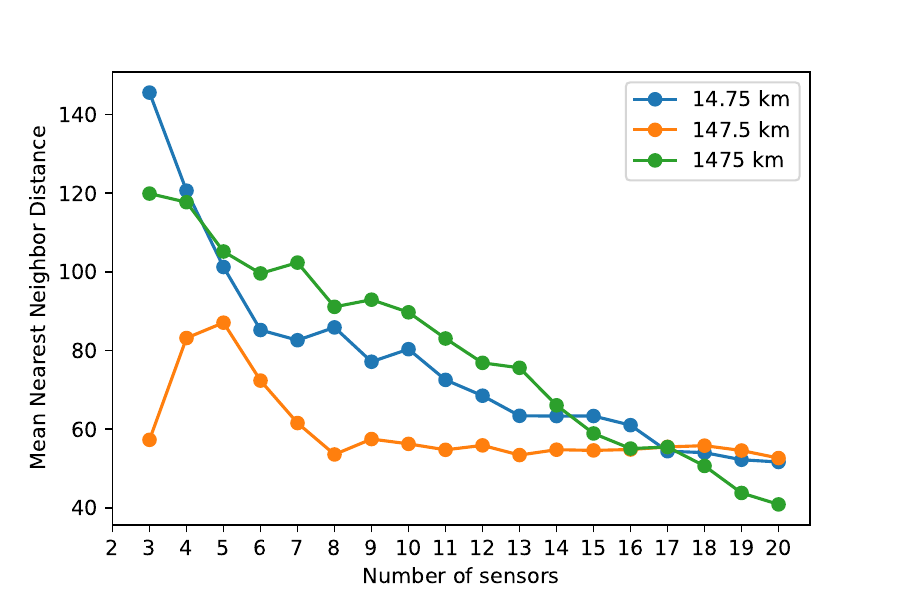}
\end{subfigure}%

\caption{Analysis of the evolution of the EIG (left) and mean distance between station (right) for three different correlation length scales. Sensors were placed using both a nonuniform and uniform prior. \\\textit{Top - Nonuniform prior:} We see that initially the disparity in EIG is small then increases with the number of sensors. However, after about 10
sensors, the disparity begins to decrease.  Few patterns can be
identified in the evolution of the geometry of the network although it may be the case that stations are initially further
apart for the low correlation model.
\\\textit{Bottom - uniform prior:} We see that, like with the nonuniform prior, the disparity in EIG across correlation length scales starts small, grows, and then shrinks again. Unlike the nonuniform prior, we see that EIG levels off sharply around 6 sensors. As with the nonuniform prior, few patterns can be discerned from the network geometry plot.}
\label{fig:corr_curves}
\end{figure*}

We investigate the effect of station correlation on the placement of sensors. We look at
three different correlation length scales, $l$: 14.75km, 147.5km, and 1475km. We do this using both a uniform prior distribution and a fault-box prior. The signal-to-measurement-noise ratio in both cases  was fixed at 0 (corresponding to a standard deviation of 1.5). Twenty stations were then placed using greedy
optimization. In Figure \ref{fig:corr_curves}, we observe that at higher correlations EIG also increases. Since higher correlation means less information, this is what we would expect to see. However, we note that this EIG gain is relatively modest, especially as more sensors are placed. It is possible that this could mean that correlation does not have a large effect on sensor placement. The mean nearest neighbor distance is also very similar for all correlation values once the number of sensors grows large.
% \begin{figure*}
% \begin{subfigure}{.5\textwidth}
% \centering
% \includegraphics[width=.95\textwidth]{figs/corr_curves.pdf}
% \end{subfigure}%
% \begin{subfigure}{.5\textwidth}
% \centering
% \includegraphics[width=.95\textwidth]{figs/corr_dists.pdf}
% \end{subfigure}

% \begin{subfigure}{\textwidth}
% \includegraphics[width=\textwidth]{figs/corr_plots2.png}
% \end{subfigure}
% \caption{The left panel illustrates the evolution of the expected information gain as
% the number of sensors increases for the three different correlation models where the
% correlation changes over two orders of magnitude. We see that initially the disparity
% in EIG is small then increases with the number of sensors. However, after about 10
% sensors, the disparity begins to decrease. The right panel investigates the geometry
% of the network by looking at the distances between stations. Few patterns can be
% identified in this plot although it may be the case that stations are initially further
% apart for the low correlation model.}
% \label{fig:corr_curves}
% \end{figure*}

\section{Conclusion}
\label{sec:conclusion}
In this work, we have demonstrated and implemented a modern framework for Bayesian optimal experimental design for analysing and optimizing a seismic monitoring network. We used this framework on a seismic source location problem with uncertainty in both the detection of seismic phases and uncertainty in the arrival time. We selected these models using data from the U.S. Transportable Array and physics-based travel time modeling with earth model uncertainty. Using these models, we capture the often-ignored influence of earth model uncertainty and station correlation on travel times. We further investigate the influence of station correlation, earth model uncertainty, and phase-arrival pick uncertainty on the sensor placement and sensitivity of the monitoring network.

Our Bayesian OED approach will enable rigorous and flexible analysis and design of monitoring networks for applications like earthquake or explosion monitoring. When evaluating a monitoring network, decision makers in high-consequence domains can trust the rigor of the Bayesian approach to provide coherent uncertainty quantification. Further, decision makers may employ Bayesian OED to assess the monitoring network’s sensitivity to different types of seismic sources and locations and therefore can certify the capabilities of the network to meet design requirements. Bayesian OED may answer other questions critical to seismic monitoring such as: how may multiphenomenology data be used to reduce uncertainty; what is the appropriate sensor fidelity or earth model resolution for estimating a QoI; and how do sensor types, number, and locations influence estimates of QoIs? 

While this work provides a meaningful first step towards analysing and optimizing monitoring networks, many simplifications were made during this exploratory study. Based on these results we have identified several follow-on directions to increase its applicability to real monitoring problems:

\begin{enumerate}
\item In this work, we used a very simple Gaussian travel time model because it enabled marginalization of origin time and handling correlation between stations. As we saw, real data are much more complex and so more complex travel time models should be explored.\\
\item Further, we may extend the correlation model to include more event characteristics. We ultimately assumed that the station correlation was independent of the event and was only a function of how far apart the stations were. Real data exhibits more complex correlation structures, such as depth dependence. Further, we assumed that the detections of each station were independent. Again, we would expect this not to be true.\\
\item We also assumed that the stations were identical. Studying a heterogeneous sensor network is much more realistic. Stations are heterogeneous both because of the use of different sensors but also based upon how the stations are installed, which could introduce different uncertainties and background noise environments. Modeling this heterogeneity also would enable us to better assess the trade off between different sensor types and installation methods.\\
\item Develop methodology that is not data driven for novel sensor placements.\\
\item Finally, we may incorporate many other sources of data into this analysis. We only considered P arrivals so other seismic phases should be studied using the same workflow and incorporated into the likelihood function. Also, infrasound sensors and seismic arrays could be included to make the analysis multi-modal by providing directional information. This would then give us the ability to explore the utility of different sensor types as we could see how the expected information gain changes as we add sensors with these different modalities. We could also then deduce the types of seismic sources different data modalities most benefit.
\end{enumerate}

\section{Acknowledgments}
Kevin Monogue participated in this research while at Stanford's Institute for Computational \& Mathematical Engineering as part of the Xplore program before joining Susquehanna International Group.
The authors would like to thank Drs. Brian Williams and Josh Carmichael from Los Alamos National Laboratory for their thoughtful and constructive feedback on the manuscript.
This Low Yield Nuclear Monitoring (LYNM) research was funded by the National Nuclear Security Administration, Defense Nuclear Nonproliferation Research and Development (NNSA DNN R\&D). The authors acknowledge important interdisciplinary collaboration with scientists and engineers from LANL, LLNL, NNSS, PNNL, and SNL.
Sandia National Laboratories is a multimission laboratory managed and operated by National Technology and Engineering
Solutions of Sandia, LLC, a wholly owned subsidiary of Honeywell
International Inc., for the U.S. Department of Energy’s National Nuclear
Security Administration under contract DE-NA0003525. This paper
describes objective technical results and analysis, which is also archived
in the Sandia report SAND2022-13022. Any subjective views
or opinions that might be expressed in the paper do not necessarily
represent the views of the U.S. Department of Energy or the United
States Government. JC contributed to the algorithm design, conducted the experiments, and wrote the manuscript under the supervision of TC. KM developed the initial statistical models. RV assisted with analysis of algorithm performance. TC managed the project, developed the theoretical framework, and wrote the initial draft of the manuscript.

\section{Data Availability}
The sensor arrival and detection data used to build the likelihood models are from the IRIS Transportable Array dataset collected between August 2007 and August 2008, available at \url{https://anf.ucsd.edu/tools/events/}.
The earth model cross sections used to build the travel time models are from the Crust 1.0 dataset, available at \url{https://igppweb.ucsd.edu/~gabi/crust1.html}. The accompanying software to this paper is hosted at \url{https://github.com/sandialabs/seismic_boed}.

\appendix
\section{Modeling Details}
\subsection{Detection Model}
 We re-emphasize that only 23\% of the event-sensor pairs contained a detection. We therefore tuned our data to balance the performance of the model. We fit the logistic regression model by minimizing binary cross entropy loss across the dataset. This loss comes from the KL divergence between the predicted detection probability and the realized detection.  Placing higher weight in the loss function on detections biases the model to predicting detection and balances the dataset composition. We experimented with different weights, Table \ref{tab:loss_weights} summarizes these results.
\begin{figure}
\begin{subfigure}[b]{.5\textwidth}
\centering
\includegraphics[width=1\textwidth]{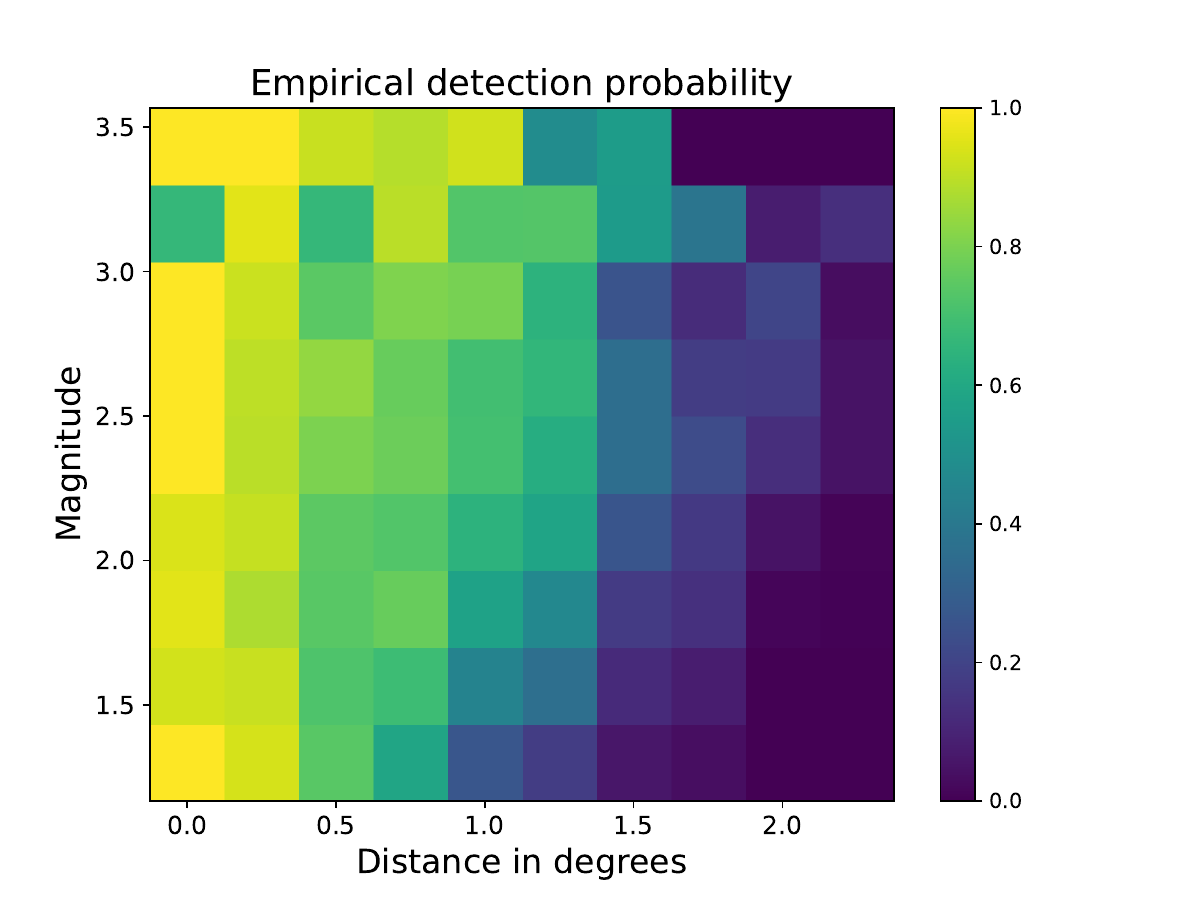}
\end{subfigure}%
\begin{subfigure}[b]{.5\textwidth}
\centering
\includegraphics[width=1\textwidth]{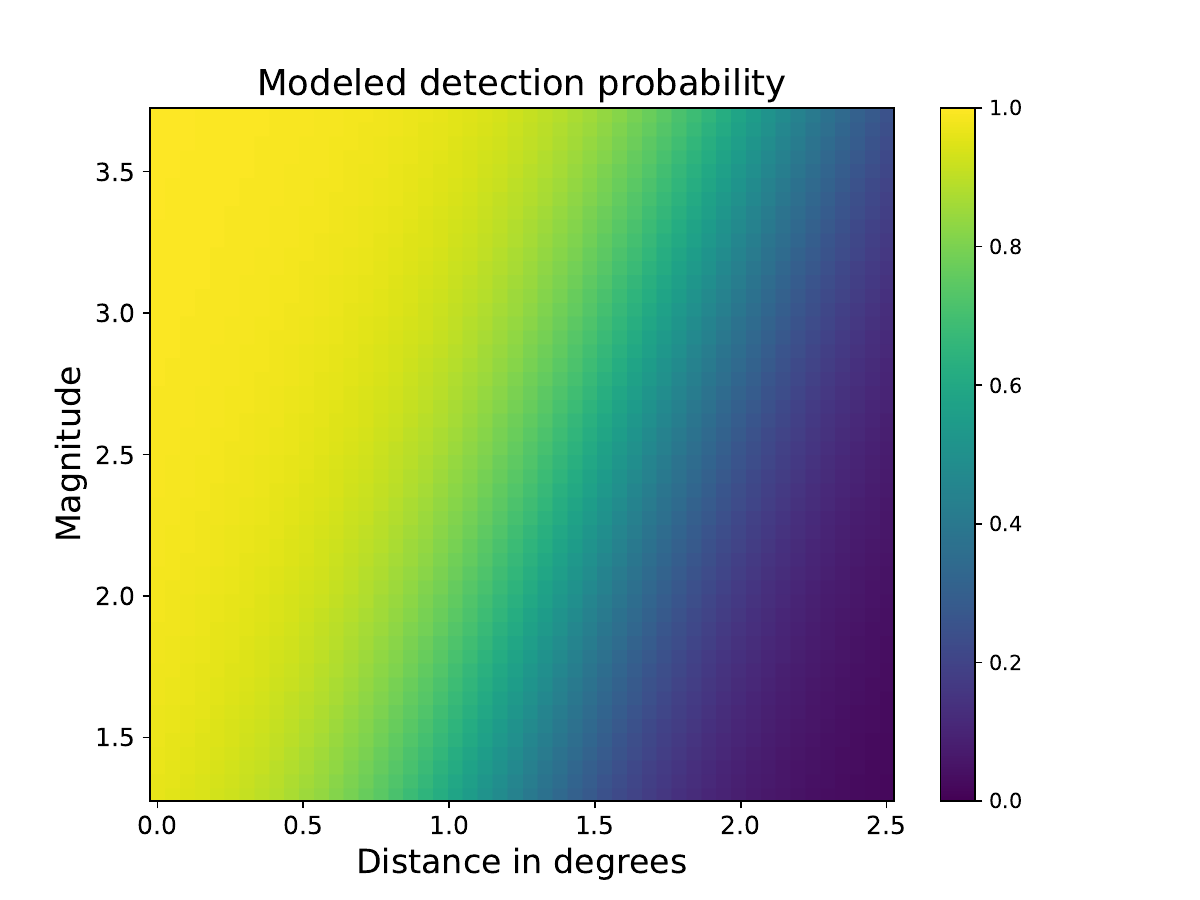}
\end{subfigure}%
\caption{The left panel shows the estimated detection probability from the Transportable Array dataset. We binned these data by distance and magnitude and then estimated the detection probability as the number of detections versus the number of potential detections for stations and events within a given distance and magnitude. The right panel shows the detection probability predicted by a logistic regression model fit to the data from the Transportable Array dataset. The modeled detection probabilities appear as a smoother version of the empirical data histogram, indicating that the model captures the underlying distribution of the data well.}
\label{fig:detect_prob}
\end{figure}

\begin{table}
	    \centering
	   % \bigskip
	    \begin{tabular}{|c|c|c|c|c|}
	    \hline
		Detection Weight  & Accuracy & Precision & Recall & AUC \\ 		\hline
		1  &  0.870   &  0.728  & 0.686 & 0.92      \\ 
		2  & 0.865   &  0.665  & 0.820 & 0.92      \\ 
		3  & 0.850   &  0.617   & 0.874 & 0.92      \\ 
		4  & 0.835   &  0.583  & 0.905 & 0.92      \\
		5  & 0.819   &  0.553  & 0.920 & 0.92      \\
		\hline
	    \end{tabular}
	    	    \caption[Detection Weighting]{Over weighting of detections in the loss function was considered to correct for the dataset imbalance towards non-detections. A weight of 2 was chosen to balance the different performance metrics.}
	    \label{tab:loss_weights}
	\end{table}

A weighting of 2 was chosen to maintain accuracy while providing a significant recall boost (catching the actual positive detections, which contribute more uniquely to the information gain).

\subsection{Arrival Time Model}
\subsubsection{Data driven model}
\begin{figure}
\centering
\includegraphics[scale=0.6]{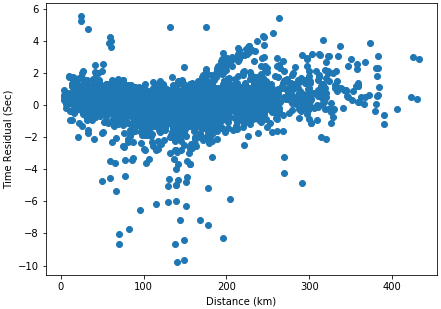}
\caption{Arrival time residual as a function of distance in kilometers.  Note that there is a bias towards positive residuals, particularly at longer distances. This bias is particularly evident in Figure \ref{fig:timeres_hist}.}
\label{fig:timeres_scatter}
\end{figure}

While the analysis in the rest of this document does not rely on a data driven arrival time likelihood model, it is helpful to consider the complexities of real arrival time data to understand some of the modeling choices. Again, the same Transportable Array dataset was used. For each P arrival, we predict the arrival time for a phase given the event and sensor locations and origin time in the catalog using the IASP91 velocity model. Then, we calculated the residuals observed in the data. Figure \ref{fig:timeres_scatter} shows a scatter plot of the residual data as a function of distance.

\begin{figure}
\centering
\includegraphics[scale=0.5]{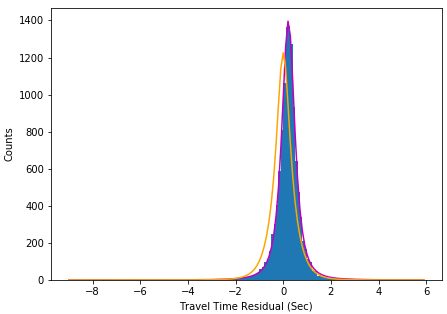}
\caption{Arrival time residual histogram compared to the fit of different statistical models. The magenta line indicates the non-centered t-distribution fit to the data. The orange line shows the marginal residual distribution under the more tractable distance dependent Gaussian model discussed in Section \ref{sec:ttu}. Note that the non-data driven model does not a priori know the bias so it is centered at zero.}
\label{fig:timeres_hist}
\end{figure}

We see little obvious relationship between distance and the residual. This residual is probably a combination of many factors: measurement noise, travel time model errors, phase categorization errors, and location errors in the catalog. In the histogram Figure \ref{fig:timeres_hist}, we see that the residuals follow a heavy tailed distribution. We choose to model it with a non-centred t-distribution.


For the non-centered t-distribution we fit the data and found that the degrees of freedom parameter was 2.198, location parameter was 0.214, and scale parameter was 0.293. While this distribution fits the data reasonably well, it does not give us the ability to tune the various sources of uncertainty when analysing and optimizing the seismic network. Further, considering station correlation and marginalizing out origin time uncertainty is very hard for this distribution. Therefore, we instead turn to a simple Gaussian distribution because the Gaussian distribution allows us to easily model correlation and marginalize out origin time uncertainty analytically. Finally, for the purpose of Bayesian OED, ignoring the bias and choosing to use a zero mean Gaussian is justified because adding a constant, known, bias to all travel times would only affect the time of the arrivals but not their uncertainty and therefore the likelihood would be the same. Obviously, for inference with real data including the bias is necessary.

\subsubsection{Improper uniform prior}
\begin{figure}
    \centering
    \includegraphics[width=\textwidth]{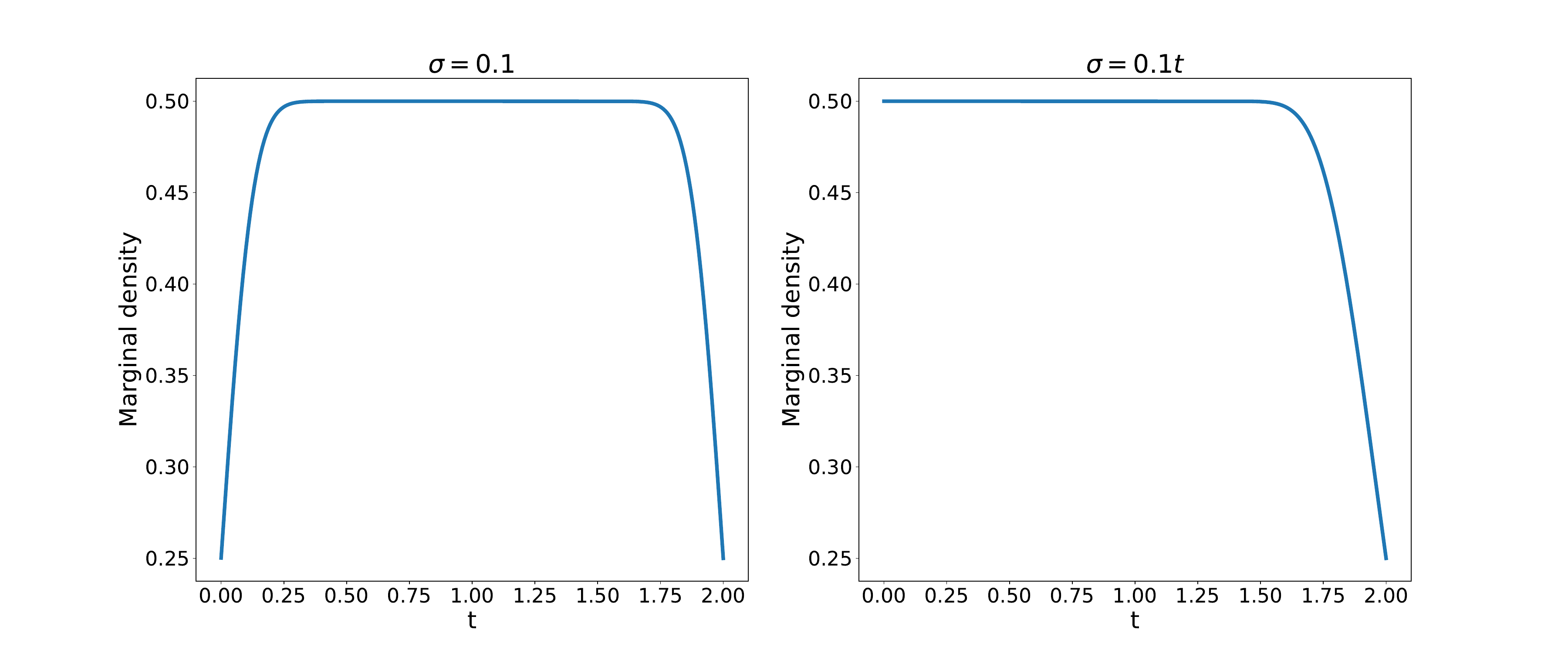}
    \caption{These plots show the one-dimensional marginal likelihood for a given travel time prediction $t$ if the origin time $t_0$ is in [$a-T, a$] (meaning the origin time is before the measured arrival time). For an improper uniform prior, this marginal likelihood should be constant on $(-\infty, \infty)$. In the plot on the left, an interval length of $T=2$ and a standard deviation of $\sigma=0.1$ are used. The marginal likelihood matches that of an improper prior except near $t=0$ and $t=T$. On the right, an interval length of $T=2$ and a standard deviation of $\sigma=0.1t$ (i.e., the error is a percentage of the mean travel time) are used. Here differences only appear near $t=T$, and the likelihood is otherwise constant.} 
    \label{fig:improper_prior}
\end{figure}

Figure \ref{fig:improper_prior} shows an empirical comparison between a marginal mean travel time likelihood using an improper prior and one using a proper prior. We see that differences between the two arise only on the boundaries of the domain of the proper prior. We note that predicted mean travel times are necessarily restricted by the size of the event domain, so when the proper uniform prior is wide enough to accommodate all possible travel times, such discrepancies will not influence the likelihoods of any observed data. In this case, the proper and improper priors are functionally equivalent.

\subsection{Importance Sampling}\label{sec:importance_sampling}
For many applications of OED, there are reasonable prior distributions from which sampling is prohibitively difficult (e.g. complex fault geometries). Further, even when sampling from a prior is easy, it might not be the most computationally efficient method for estimating the integrals in \eqref{eq:thetaEIG} because events that are rare according to the prior may contribute significantly to the integral (e.g. high magnitude seismic events that are likely to cause very high information gains). Thus, to be able to fully utilize domain knowledge about areas of interest in an efficient manner it is important that we have a way to representatively sample from these challenging priors and important events. Importance sampling is one such way.

We draw samples from an importance distribution ($q\left( \theta^\prime\right)$, a distribution different from the prior but one that is possible to sample), weight those samples according to the probability density function of our target prior distribution, $p\left( \theta^\prime\right)$, and use these weighted samples to approximate our quantities of interest. This means that in Algorithm 1 instead of $\theta^\prime \sim p\left( \theta^\prime\right)$, we have that $\theta^\prime \sim q\left( \theta^\prime\right)$ and has a corresponding weight of $w\left (\theta^\prime \right) = p\left( \theta^\prime\right)/q\left( \theta^\prime \right)$. For a detailed discussion of importance sampling see \cite{ref:mcbook}.

\subsection{Description of Software Implementation}\label{sec:appdx_software}
The accompanying software to this paper is hosted at \url{https://github.com/sandialabs/seismic_boed}.
The user can specify models for generating synthetic data and assessing the likelihood of that synthetic data for different sensors and events in the domain of candidate events. The code is separated into two main components: analysis network and optimization network. The code is designed to use MPI so that it can run on HPC resources. The multi-core parallelism through MPI is implemented by \textsc{mpi4py} \citep{dalcin2021mpi4py}. The EIG computational is highly parallelizable so it can be scaled easily to thousands of cores, which is important due to the number of computations required for robust estimates of the EIG, particularly when making the sensitivity maps to show how the network performs on specific events.

The analysis code estimates the EIG of a given seismic monitoring networks for a user-defined prior distribution of potential events. As described in Algorithm 1, the code samples these candidate events and then generates synthetic datasets that could plausibly be seen by the sensors. Likelihood models for several sensor types are provided but user-specified models can also be used. For each of the datasets the code constructs the posterior distribution and computes the information gain IG according to the KL-divergence. This information gain is averaged over all synthetic datasets to compute the EIG. The code can also return a list of the IG for different hypothetical events which can be used to generate a map of sensitivities of the network to different event locations, depths, and magnitudes. See Figure \ref{fig:single_heatmap} for an example. 
        
The optimization code is a wrapper around the analysis code. Given a specified initial network configuration of sensors, the code will add a desired number of sensors to the network. The goal of the optimization is to maximized the EIG of the new sensor network while respecting user-specified constraints on where sensors can be placed. This is done with a sequential (greedy) optimization that adds sensors one at a time to the initial network. Each optimization is done using a Bayesian optimization method that construct a Gaussian process surrogate model of the EIG optimization surface. This is done by evaluating many potential new sensor locations and measuring the EIG using the analysis code. These data are then used to construct the surrogate and inform new trial points to query the EIG function. The code then returns the new sensor network after the optimal sensors have been added.

Please refer to the documentation in \cite{catanach2024github} for a complete description of the code, capabilities, and provided tutorials.
\subsection{More results}
\begin{figure}
    \centering
     \centerline{\includegraphics[width=1\linewidth]{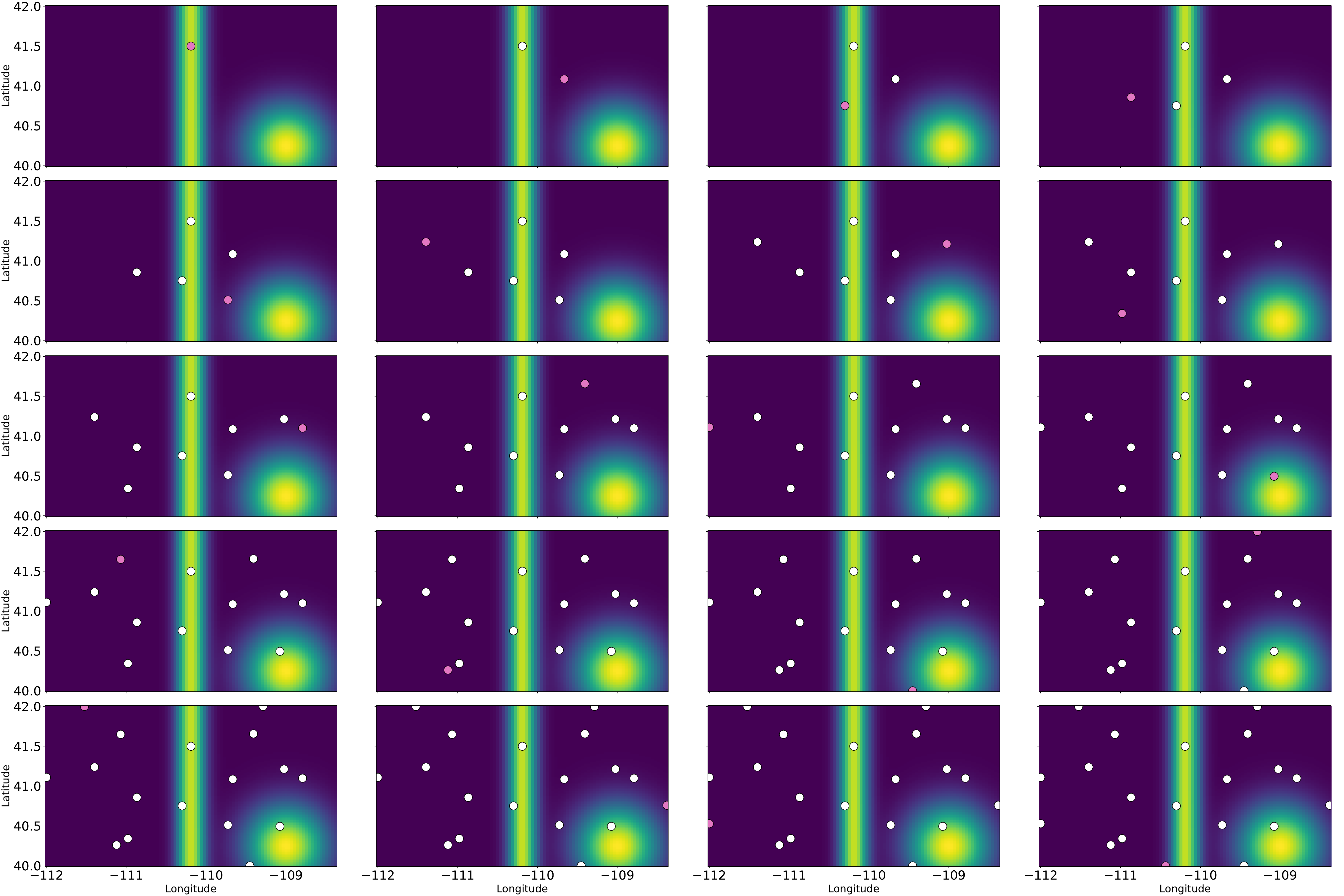}}
\caption{Illustration of the sequential placement of sensors on a square domain when using a nonuniform prior on events. Each plot shows the sensors placed in previous steps in white and the sensor placed at the current step in pink. The location component of the prior is depicted by the heat map underneath the sensors, with warmer areas indicating areas of higher prior density. We see that the sensors are generally placed in areas of lower prior density.}
\label{fig:sequential_placement}
\end{figure}

\begin{figure}
    \centering
    \centerline{\includegraphics[width=1\linewidth]{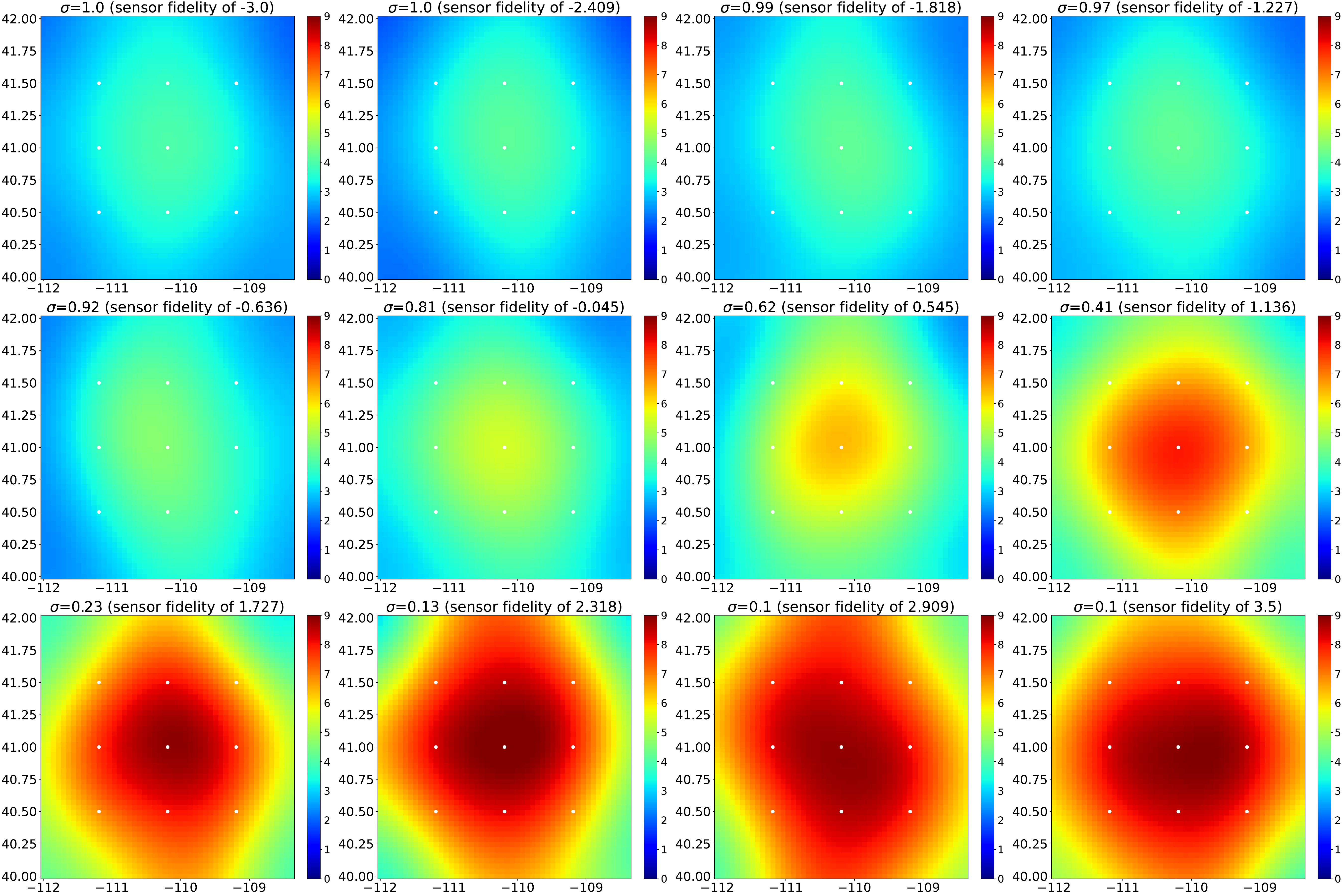}}
    
\caption{Illustration of the effect of changing the measurement uncertainty standard
deviation over several orders of magnitude when using a uniform distribution. The color plots illustrate the EIG of
a shallow seismic source with the different stated measurement errors. 
For noise levels below about 0.41 seconds the model uncertainty dominates over measurement
uncertainty so EIG is fairly stable.}
\label{fig:fid_surface_uniform}
\end{figure}

\begin{figure}
      \centerline{ \includegraphics[width=1\linewidth]{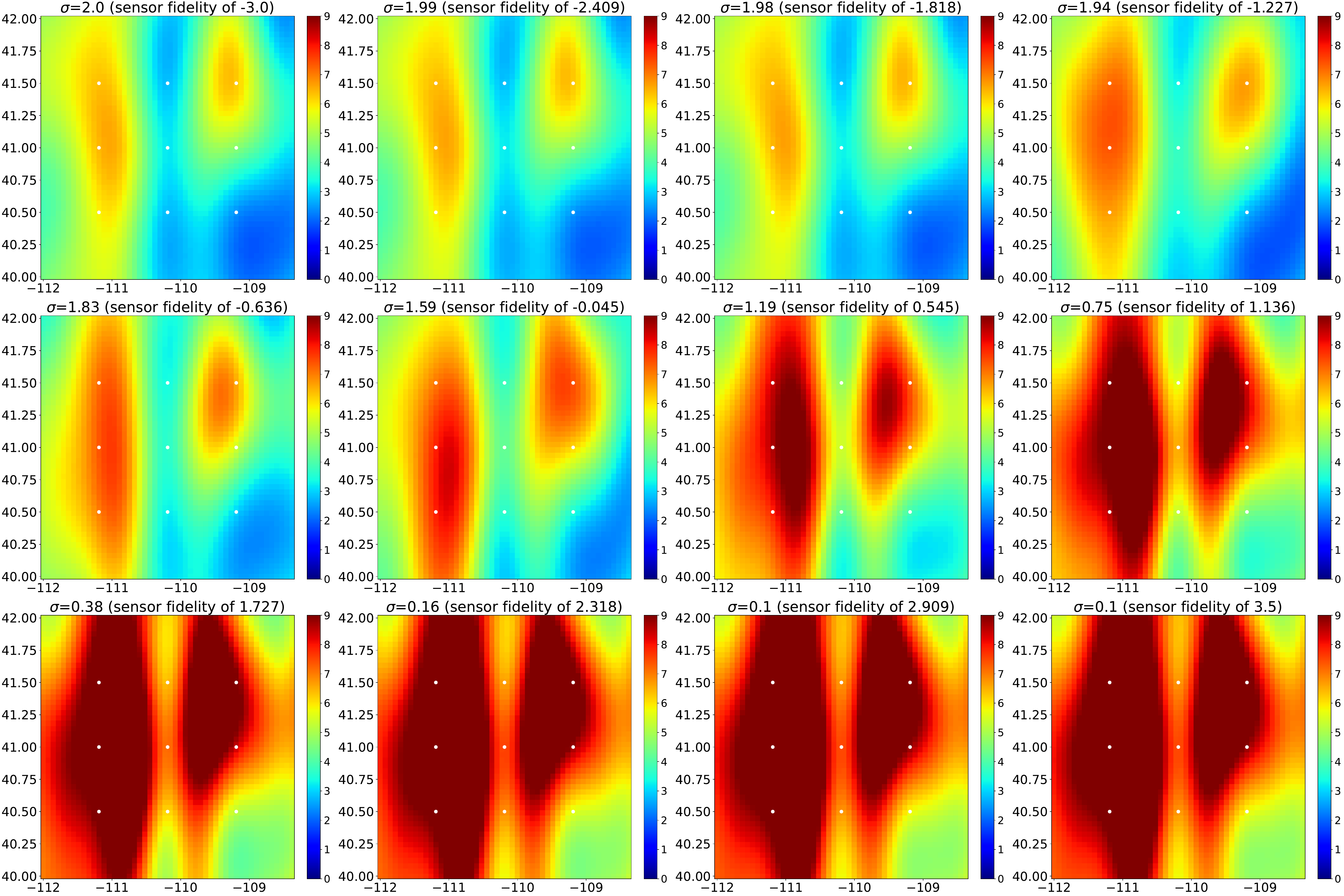}}
    \caption{Illustration of the effect of changing the measurement uncertainty standard
deviation over several orders of magnitude when using a non-uniform distribution. The color plots illustrate the EIG of
a shallow seismic source with the different stated measurement errors. 
For noise levels below about 1.59 seconds the model uncertainty dominates over measurement
uncertainty so EIG is fairly stable.}
\label{fig:fid_surface_faultbox}f
\end{figure}

This section contains additional figures illustrating various results of the paper. Figure \ref{fig:sequential_placement} demonstrates the process of sequentially placing sensors using a nonuniform prior. Figures \ref{fig:fid_surface_uniform} and \ref{fig:fid_surface_faultbox} show the effect of changing the sensor fidelity on Expected Information Gain.

\clearpage

\bibliographystyle{unsrtnat}
\bibliography{bib}

\end{document}